\documentclass[aos]{imsart}
\RequirePackage{amsthm,amsmath,amsfonts,amssymb}
\RequirePackage[authoryear]{natbib}
\RequirePackage[colorlinks,citecolor=blue,urlcolor=blue]{hyperref}
\RequirePackage{graphicx}

\usepackage{enumerate}
\usepackage{url} 
\usepackage{float}
\usepackage{tcolorbox}
\usepackage{amssymb}
\usepackage[official]{eurosym}
\usepackage{longtable}
\newtheorem{remark}{Remark}
\usepackage{subcaption}
\usepackage{adjustbox}
\usepackage{rotating} 
 \usepackage{soul}
\usepackage{etoc}
\usepackage[ruled,vlined]{algorithm2e}
\usepackage{xcolor}
\usepackage{multirow}  
\usepackage{adjustbox}
\usepackage{amsthm}
\newcommand{\LJ}{\textcolor{black}}
\startlocaldefs

\newcommand{\E}{\mathbb{E}}
\newcommand{\p}{\mathbb{P}}

\newcommand{\R}{\mathbb{R}}

\newcommand{\bs}{\boldsymbol}
\newcommand{\FF}{\mathcal F}
\newcommand{\mf}{\mathbf}

\newcommand{\by}{\mathbf{y}}

\newcommand{\bepsilon}{\bs{\epsilon}}

\newif\ifarxiv
\arxivtrue   
\ifarxiv
  \newcommand{\apxref}[1]{Supplementary Material}
\else
  \newcommand{\apxref}[1]{Appendix \ref*{#1}}
\fi

\usepackage{titlesec}
\titlespacing\section{0pt}{0pt plus 4pt minus 2pt}{0pt plus 2pt minus 2pt}
\titlespacing\subsection{0pt}{0pt plus 4pt minus 2pt}{0pt plus 2pt minus 2pt}
\titlespacing\subsubsection{0pt}{0pt plus 4pt minus 2pt}{0pt plus 2pt minus 2pt}

\newtheorem{assumption}{Assumption}[section]

\newtheorem{theorem}{Theorem}

\endlocaldefs

\begin{document}

\begin{frontmatter}
\title{Complex trend inference for high-dimensional piecewise locally stationary time series} 
\runtitle{AJDN-H}

\begin{aug}
\author[A]{\fnms{Lujia}~\snm{ Bai}\ead[label=e1]{lujia.bai@rub.de}},
\author[B]{\fnms{David}~\snm{Veitch}\ead[label=e2]{david.veitch@mail.utoronto.ca}}
\author[C]{\fnms{Weichi}~\snm{Wu}\ead[label=e3]{wuweichi@mail.tsinghua.edu.cn}}
\author[D]{\fnms{Wenyang}~\snm{Zhang}\ead[label=e4]{wenyangzhang@um.edu.mo}}
\and
\author[B]{\fnms{Zhou}~\snm{Zhou}\ead[label=e5]{ zhou.zhou@utoronto.ca}}
\address[A]{ Fakult\"at f\"ur Mathematik, Ruhr-Universit\"at Bochum\printead[presep={,\ }]{e1}}

\address[B]{Department of Statistical Sciences, University of Toronto\printead[presep={,\ }]{e2,e5}}
\address[C]{Department of Statistics and Data Science, Tsinghua University\printead[presep={,\ }]{e3}}
\address[D]{Faculty of Business Administration and APAEM, University of Macau\printead[presep={,\ }]{e4}} 

\end{aug}

\begin{abstract}

This paper studies high-dimensional trend inference for piecewise smooth signals under nonstationary noise and asynchronous structural breaks by first detecting asynchronous changes without assuming stationarity and then further exploiting latent group structures to estimate trend functions.
In the first step, we propose AJDN (Asynchronous Jump Detection under Nonstationary Noise), a multiscale framework for the identification and localization of jumps in high-dimensional time series. We show that AJDN consistently recovers the number of jumps with a prescribed asymptotic probability and achieves nearly optimal localization rates in the presence of asynchronicity and nonstationarity, both of which often violate the assumptions of existing high-dimensional change point  methods and thereby deteriorate their performance. 
In the second step, we augment AJDN with a homogeneity pursuit step and obtain AJDN-H, 
which identifies latent groups of dimensions that share common jump structures and trend parameters given the detected jumps. This allows for efficient information pooling and improves the accuracy of trend estimation under both asynchronicity and nonstationarity. The robustness and finite-sample performance of the proposed methodology are examined by extensive simulation studies. 
An application to financial data demonstrates the practical utility of the AJDN-H framework in complex, high-dimensional settings.
\end{abstract}

\begin{keyword}[class=MSC]
\kwd[Primary ]{	62G05 }
\kwd{62G10}
\kwd[; secondary ]{62M10}
\end{keyword}

\begin{keyword}
\kwd{High-dimensional time series}
\kwd{Nonstationarity}
\kwd{Jump detection}
\kwd{Homogeneity Pursuit}
\end{keyword}

\end{frontmatter}

\begin{figure}
    \centering
    \includegraphics[width=0.9\linewidth]{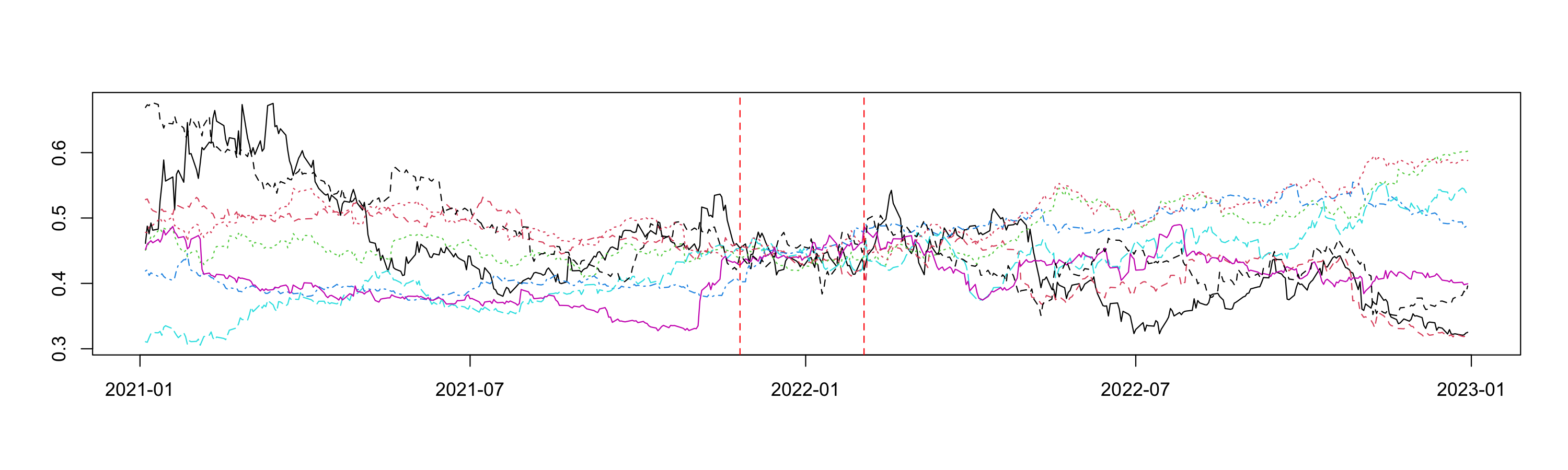}
    \caption{Trajactories of 8 of Nasdaq-100 stocks ("ABNB" "AMZN", "ADI", "AAPL", "ODFL", "QCOM", "TTWO", "TXN"). The red dotted lines are 6-th and 7-th detected unique jumps by our method, between which the 8 stocks share the same mean structure}
    \label{fig:raw}
\end{figure}

\section{Introduction}


\ \ 
In high-dimensional time series, there are three complications that make the inference of piecewise smooth trends 
particularly challenging. First, in many real applications, the noise process in at least one dimension exhibits nonstationary behavior. Such complex temporal dynamics have been documented in the physical sciences, for example in climatology (see the summary in \cite{zhou2013heteroscedasticity}), as well as in economics and finance \citep{hardy2001regime,li2019time}, among many other areas. Second, the trends are complex in the sens of exhibiting structural breaks that could occur at different time points across dimensions, even when the underlying source of the changes is shared. Third, the shape of the high-dimensional trends  
may share the same structure across different sub-periods and coordinate. 
In Figure \ref{fig:raw}, we show the 8 stocks from Nasdaq-100. We divide each  closing daily stock price adjusted for dividends by the value of the price of QQQ, an exchange-traded fund which tracks the Nasdaq index. Although each stock has its  own dynamics and jump points in mean, they share similar mean behaviors around the January of 2022. 

To date, inference for trends in high-dimensional time series, including jump detection and estimation, has largely overlooked these three issues. As a result, existing methods may suffer from reduced sensitivity, unreliable false positive control, or less accurate trend estimation in applications. The aim of this article is to develop a unified two-step framework for inference on trend functions under nonstationary noise with both temporal and cross-sectional dependence. In the first step, we study the detection and localization of possibly asynchronous jumps via  novel multiscale local test statistics together with a simple multiplier bootstrap scheme to approximate their asymptotic behavior. In the second step, we investigate the refined estimation of trend functions which leverages both smoothing and the common structure identified across component series over certain time intervals by a new clustering approach based on differences between the local linear fits of trends. 

\subsection{Problem setup.}\  Assume the $p$-dimensional time-verying trend model 
\begin{align}
    \mf Y_i = \bs \beta(i/n) + \bs \epsilon_i, \quad i =  1, \ldots, n, \label{eqn:mainmodel}
\end{align}
where $\mf Y_i = (y_{1,i},\cdots, y_{p,i})^{\top}$, 
and $\boldsymbol{\beta}(t)=\left(\beta_1(t),\dots,\beta_p(t)\right)^\top$ is a $p$-dimensional piecewise smooth mean function where each $\beta_r(t)$ may or may not experience jumps, and $\boldsymbol{\epsilon}_i$ is a $p$-dimensional centred non-stationary error process which can evolve both smoothly and abruptly throughout time. Further details on the mean function and the non-stationary error process can be found in Sections \ref{sec:assumptionjumps} and \ref{sec:apls}, respectively.  Suppose in each dimension $1 \leq r \leq p$, there are $m_{r,n}$ jumps $0=d_{r,0} < d_{r,1} < \cdots <  d_{r,m_{r, n}+1} = 1$. Altogether, there are $m_n=\sum_{i=1}^rm_{i,n}$ unique jumps (after combining the ones that are close to each other), $0=d_{0} < d_{1} < \cdots <  d_{m_n+1} = 1$. Define $\mathcal I_{r, q} = (d_{r,q}, d_{r, q+1}]$, $\mathcal I_{l} = (d_l, d_{l+1}]$. 
We assume that on each $\mathcal I_{r, q}$, $\beta_{r}(t)$ poses Lipschitz continuous second-order derivatives. On each $\mathcal I_{l}$, we consider the case where $\bs \beta(t)$ admits a spatial homogeneity structure with a fixed cardinality of $K$, i.e., there exists a partition of the $\{1,2,\ldots, p\}$ denoted by $\{ C_{1,l}, \ldots,  C_{K_l^0,l}\} $, $C_{i,l} \cap C_{j,l} = \emptyset$, $1 \leq i \neq j \leq K_l^0$,
\begin{align}
    \beta_{r}(t) = \alpha_{k,l}(t), \quad r \in C_{k,l},~ t \in \mathcal I_l, ~1 \leq k \leq K_l^0,~1 \leq l \leq m_n+1,
\end{align}
where $K_l^0$ is the number of clusters in the $l$-th piece of time series vectors. The area between the two dotted lines in Figure \ref{fig:raw} belongs to the same cluster in that piece of time from 2021-11-26 to 2022-02-02.

To infer the trends, a set of three questions naturally arises. (a) Can we test whether there are changes? (b) Can we locate these changes? (c) Can we identify the common group structures and exploit it to improve estimation by pooling information across components?
In this paper, we study these three questions for the high-dimensional piecewise locally stationary time series, allowing for both changes and jumps in the mean as well as break points in the high-dimensional error process.
This framework is particularly relevant for financial data, which often exhibits local stationarity in practice; see, for example, \cite{sundararajan2018nonparametric, ng2026inference}. Our methodology provides a statistical foundation for detecting clusters of assets that share common regime behavior, thereby offering insight into time-varying market co-movements and the propagation of systemic risk, phenomena widely documented in the literature; see, for example, \cite{ABDOLLAHI2024102004}. See also the data analysis in Section \ref{sec:stocksanalysis}.
To address questions (a) and (b), we introduce AJDN, and to address question (c), we further develop AJDN-H.

\subsection{Detecting jumps with AJDN. } \ 
For (a) and (b), we propose the AJDN  method
which applies $W(\cdot)$, an optimal jump-pass filter \citep{wu2019multiscale}, at various times $t$ and scales $s$ to each individual dimension $r$ of the high-dimensional time series
\begin{align}
	H(t,s,r)&=\frac{1}{\sqrt{ns}}\sum_{j=1}^ny_{r,j} W\left(\frac{j/n-t}{s}\right) \label{eqn:htsr}\\
    G_{\text{max}}(\mathbf{T})&=\sup_{\substack{1\leq r\leq p \\ \underline{s}_r\leq s\leq \bar{s}_r \\ t\in T_r}} G(t,s,r)=\sup_{\substack{1\leq r\leq p \\ \underline{s}_r\leq s\leq \bar{s}_r\\ t\in T_r}} \frac{|H(t,s,r)|}{\sqrt{\hat{\sigma}_{r,t}^2}}. \label{eqn:firstmaxstatistic}
\end{align}

\eqref{eqn:firstmaxstatistic} is the first stage AJDN test statistics, and the corresponding further information, including the variance adjustment factor $\hat{\sigma}^2_{r,t}$, can be found in Section \ref{sec:methodologyoverview}. Notably, \eqref{eqn:firstmaxstatistic} does not pool evidence that a jump occurs at a specific time across dimensions. After $H(t,s,r)$ is calculated for all times, scales, and dimensions, a maximum over all normalized values of these statistics is taken as $G_\text{max}(\mathbf{T})$, and the critical value of $G_\text{max}(\mathbf{T})$ at a desired significance level is estimated via a high-dimensional block multiplier bootstrap scheme.  At the second stage of AJDN,   a high-dimensional local CUSUM procedure is employed to further improve the accuracy of the localization of jumps to an asymptotically nearly optimal rate under the asynchronous jump assumption.


Moreover, AJDN is able to address the first two complications frequently encountered in high-dimensional jump detection raised at the beginning of the paper. First, our simple multiplier bootstrap scheme is able to mimic the covariance structure of the high-dimensional vector of $H(t,s,r)$ test statistics, enabling us to estimate critical values of the maximum of this vector of test statistics even in the presence of nonstationary errors. Second, the construction of AJDN's test statistic and algorithm for detecting multiple jumps does not pool evidence across dimensions and makes no assumption on the minimum spacing between jumps in different dimensions, which allows AJDN to detect jumps that occur either synchronously or asynchronously across multiple dimensions. As well, an additional benefit of our test is its multiscale nature frees users from making a consequential decision of what scale is optimal for jump detection, which can be sensitive to the mean's evolution between jumps. In the high-dimensional setting this is a particularly pertinent issue as the form of the trend function can vary substantially between dimensions.


\subsection{AJDN-H : Homogeneity pursuit.} \ 
Based on AJDN, to answer question (c) and address the third complication at the beginning of the paper, we propose identifying the common structures by extending hierarchical clustering to the non-parametrically estimated mean functions. 
A key point to pruning the clustering trees is to extend classical information criteria such as AIC (Akaike, 1974) and BIC (Schwarz, 1978) to high-dimensional, nonstationary frameworks, which is highly non-trivial. In particular, we are interested in the information criterion of the form 
 \begin{align}
        \mathrm{IC}(K, l) = \log (\tilde \sigma_{l}^2(K) )+ K \chi_n, \label{eq:ICintro}
    \end{align}
    where $K$ is the number of clusters, $\chi_n$ is the penalty coefficient, $\tilde \sigma_{l}^2(K)$ is the normalized sum of squares of non-parametric residuals for the $l$-th piece of high-dimensional time series vectors.
    
  There are several challenges in using \eqref{eq:ICintro}. First, the effective sample size within each locally stationary segment is often small relative to full sample size, resulting in a nonstandard asymptotic regime in which the normalizing rate for the sum of squares of non-parametric residuals must be chosen to achieve a nonzero limit. Second, in piecewise locally stationary settings, both the number of parameters and the segment lengths  must be estimated from the data, which introduces additional estimation uncertainty and complicates the asymptotic expansion of the information criterion \citep{cho2015, wang2021optimal}. Finally, conventional penalty terms of order $O(\log n)$ may be insufficient to control overfitting, since the nonparametric residual term in $\tilde \sigma_l^2(K)$ typically converges much more slowly than its parametric counterpart.

\subsection{Literature review.} 
\ 

As the amount and variety of data being collected across disciplines grow, the importance of high-dimensional change point detection methods is increasingly pronounced, with recent literature reviews including \cite{liu2022high}. However, most methods proposed to date are not robust to error processes with complex temporal dynamics. 
A number of high-dimensional change point detection methods have been proposed under the assumption of increasingly complex errors, such as i.i.d normal errors \citep{enikeeva2019,grundy2020high}, i.i.d. errors \citep{2022wangvolgushev,zhang2022adaptive}, and stationary errors \citep{jirak2015uniform,2018dette, chen2021inference,li2022ell}, however we are unaware of any work in the high-dimensional setting which relaxes the assumption further and allows for nonstationary errors. Moreover, asynchronous jumps, where different dimensions experience jumps at slightly different times, pose a challenge for many existing high-dimensional change point methods because they violate the minimum-spacing assumptions imposed by these methods. As a result, such methods may be unable to detect jumps occurring close to previously detected ones, or may terminate once the tested segment becomes too short. In contrast, AJDN-H assumes a minimum spacing between jumps in the same dimension. However, it differentiates itself from previous works in that jumps in different dimensions may occur arbitrarily close to one another. 


Existing methods for nonparametric trend estimation in high-dimensional time series with a diverging number of jumps often suffer from boundary effects, leading to less accurate estimates near structural breaks, see \cite{bai2025}. To address this issue, we propose incorporating homogeneity pursuit, which leverages cross-sectional similarity to stabilize estimation and mitigate boundary distortions. Homogeneity pursuit aims to group individuals with identical but unknown parameters, thereby identifying those samples that provide information relevant to the attributes of a given target individual.  A growing body of literature over the past decade has investigated homogeneity pursuit via penalization techniques (see \cite{ke2015homogeneity, Kong2017, Tian02102023, sun2024multi} among others), binary segmentation strategies (see \cite{ke2016} among others) . Recently, clustering-based methods have attracted handsome attention, see \cite{VOGT2020305, han2025new, sun2025homogeneity}. However, the aforementioned literature mostly focuses on identifying common groups for the entire interval and  there has been very limited literature discussing homogeneity pursuit in the presence of jumps and local stationarity.

The rest of this paper is organized as follows. Section \ref{sec:methodologyoverview} provides an overview of the proposed AJDN method to detect jumps, identify common groups and estimate mean functions. Section \ref{sec:assumptions} describes the assumptions posited upon both the mean and error processes, as well as the nonlinear filters and scales used by AJDN for the jump detection and estimation. Section \ref{sec:asympresults} details the theoretical results justification of AJDN. 
The consistency of homogeneity pursuit based on the jump estimation is discussed in Section \ref{sec:group}. In Section \ref{sec:simulationstudy} we present a simulation study demonstrating AJDN's strong power and Type I error control for change point detection relative to other competing methods. After assessing the finite sample performance of estimating the change points, we compare the empirical performance of AJDN-H in the case of homogeneity and without homogeneous structures with that only using AJDN in Section \ref{sec:simgrou}. 
In Section \ref{sec:stocksanalysis} we use AJDN-H to analyze a real  dataset from Nasdaq-100 stocks.
Details of tuning parameter selection, further numerical experiments,  the proofs of all theoretical results and additional results for the data analysis are relegated to the online supplementary material.

\section{Methodology overview.}\label{sec:methodologyoverview}

\subsection{First stage jump detection of AJDN}.\label{sec:firststagejump}
\ Given a $p$-dimensional time series $\{\mathbf{Y}_i\}_{i=1}^n = \{(y_{1,i} , \dots, \allowbreak y_{p,i})^\top\}_{i=1}^n$ where $\mathbf{Y}_i$ follows model \eqref{eqn:mainmodel}, the goal of AJDN is to detect the times and dimensions where the piecewise smooth trend experiences a jump. Recall from \eqref{eqn:firstmaxstatistic} that the first stage test statistic of AJDN is a multiscale test statistic. 
Denote $\underline{s}_r,\bar{s}_r$ as the minimum and maximum scales used in each dimension $r$ (the scales used do not have to be the same for every dimension), and correspondingly $\bar s_{\max}=\max_r \bar s_r$, $\bar s_{\min}=\min_r \bar s_r$. Section \ref{sec:assumptiononscale} discusses the selection of scales from a theoretical perspective. 
\apxref{apx:selecthyper} discusses practical issues around selecting the maximum and minimum scales to use. Once $\underline{s}_r,\bar{s}_r$ are selected for $r=1,\dots,p$, a sparse sequence of scales $\underline{s}_r=s_{r,1}\leq s_{r,2}\leq \cdots \leq s_{r,\delta_n}=\bar{s}_r$ (without loss of generality we always assume $\underline s_r<\bar s_r$ in this paper) are calculated and used in AJDN
\begin{align}\label{sparsescale}
	s_{r,i}&=2^{g_{r,i}}\\
	g_{r,i}&=\log_2\underline{s}_r+(i-1)\frac{\log_2\bar{s}_r-\log_2\underline{s}_r}{\delta_n-1},\; 1\leq i\leq \delta_n\\
	\delta_n&=C (\log n)^{1+\epsilon}\log^{5/2} (pn), \epsilon>1/2.\label{sparsescaledelta}
\end{align}
In practice, we suggest $\epsilon=.51$ and determine $C$ by the available computational resources. 
It will be shown in Section \ref{sec:sd} that the sparsified statistic is asymptotically equivalent to its continuous version \eqref{eqn:firstmaxstatistic}. AJDN's test statistic using this sparse sequence of scales and discrete times is
\begin{align}
     G_{\text{max}}^{\delta_n}(\mathbf{T})=\sup_{\substack{1\leq r\leq p \\ 1\leq j \leq \delta_n  \\ t_i\in T_r}} G(t_i,s_{r,j},r)=\sup_{\substack{1\leq r\leq p \\ 1\leq j\leq \delta_n\\ t_i\in T_r}} \frac{|H(t_i,s_{r,j},r)|}{\sqrt{\hat{\sigma}_{r,t}^2}} \label{eqn:mjpdhdmaxstat}. 
\end{align}

\begin{figure}[t]
	\centering
	\includegraphics[width=14cm]{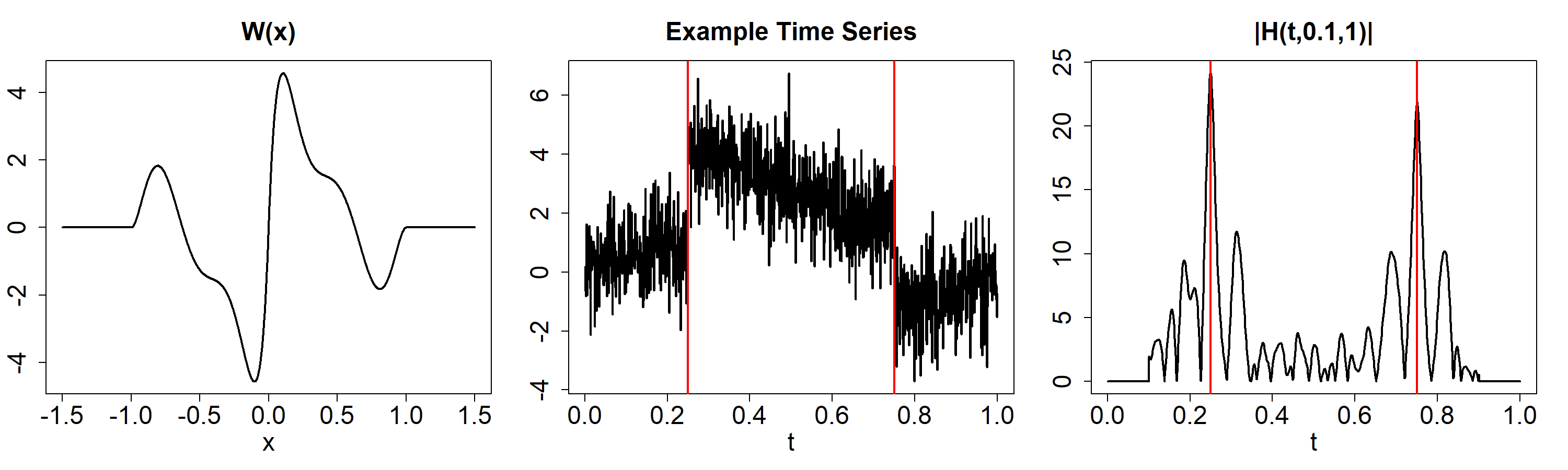}
	\caption{\it Application of optimal jump pass filter $W(x)$ to an example univariate time series with a piecewise smooth mean function, with the two jumps denoted by red lines. For a single scale $s=0.1$ we can see $|H(t,0.1,1)|$ is maximized at the jump points.}
	\label{fig:filtergraph}
\end{figure}

In the above, $\mathbf{T}$ is the set of all times, in all dimensions that a change point could be detected, specifically $\mathbf{T}=(T_1,\dots,T_p)^\top,\; T_r=(\bar{s}_r,1-\bar{s}_r]$ and $t_i=i/n$ for $i= 1,\dots,n$. Recall from \eqref{eqn:htsr} that AJDN-H involves a nonlinear filter $W(\cdot)$ which filters out noises and smooth changes in the mean and keeps the jumps; See Section \ref{sec:nonlinearfilter} for detailed assumptions on $W(\cdot)$. In this article we adopt the optimal jump-pass filter proposed in \cite{wu2019multiscale}
\begin{align}
    W(x)=\big(&-1294.2222x^6+4246.6667|x|^5-5320x^4 \nonumber\\
              &+3188.8889|x|^3-933.3333x^2+112|x|\big)\text{sign}(x)
\end{align}
which can be seen in Figure \ref{fig:filtergraph}.

AJDN's test statistic takes a maximum over a large number of local (in time and dimension) tests for the presence of a jump. An example is provided in Figure \ref{fig:filtergraph} showing that the absolute value of $H(t,s,r)$ will be large, and  locally maximized at $t$ when $t$ coincides with a jump in dimension $r$. We can also see that there is some ``leakage'' in that $H(t,s,r)$ can also take large values close to, but not exactly located at, a jump. However, given the max-type nature of our test statistic, and our algorithm for detecting multiple jumps, these large values are of little concern as conditional on the jump being detected they will be ultimately ignored by AJDN. 

For purposes of comparability across time and dimensions (there is no assumption that variance is homogenous cross-sectionally) $H(t,s,r)$ is normalized by a local estimate of the standard deviation of the time series in dimension $r$ at time $t$, using observations where $i/n \in [t-\bar{s}_r,t-\underline{s}_r]\cup [t+\underline{s}_r,t+\bar{s}_r]$: 
\begin{align}
    \hat{\sigma}_{r,t}^2&:=\frac{\left(\sum_{i \in K_{r,\text{left}}(t)}(y_{r,i}-\bar{y}_{r,\text{left}}(t))^2 + \sum_{i\in K_{r,\text{right}}(t)}(y_{r,i}-\bar{y}_{r,\text{right}}(t))^2\right)}{|K_{r,\text{left}}(t)|+|K_{r,\text{right}}(t)|} ,\\
	\bar{y}_{r,\text{left}}(t)&=\frac{1}{|K_{r,\text{left}}(t)|}\sum_{i \in K_{r,\text{left}}(t)} y_{r,i}, \;\;\;
	\bar{y}_{r,\text{right}}(t)=\frac{1}{|K_{r,\text{right}}(t)|}\sum_{i \in K_{r,\text{right}}(t)} y_{r,i},	\\
 K_{r,\text{left}}(t)&= \{i\;|\; i/n \in [t-\bar{s}_r,t-\underline{s}_r]\},\;\;\;K_{r,\text{right}}(t)= \{i\;|\; i/n \in [t+\underline{s}_r,t+\bar{s}_r]\},\\
	|K_{r,\cdot} (t)|&= \text{Cardinality of } K_{r,\cdot}(t).
\end{align}

Given the distribution of AJDN's test statistic $G^{\delta_n}_{\text{max}}(\mathbf{T})$ could be very complicated and unlikely to be known, especially for a non-stationary noise sequence, a high-dimensional block multiplier bootstrap procedure is used to estimate the $(1-\alpha)$ quantile of $G^{\delta_n}_{\text{max}}(\mathbf{T})$. Specifically, for a given block size parameter $s'\leq \min_{1\leq r\leq p}\underline{s}_r$,
\begingroup
\allowdisplaybreaks
\begin{align}
    \boldsymbol{\Upsilon}_i&=\frac{1}{\sqrt{2ns'}}\Bigg(\sum_{j \in [i-s'n,i)} \mathbf{Y}_j-\sum_{j \in [i,i+s'n)} \mathbf{Y}_j\Bigg) \label{eqn:upsilon}\\
	&=(\Upsilon_{1,i},\dots,\Upsilon_{p,i})^\top\\
	\hat{G}^{(\ell)}_{\text{max}}(\mathbf{T})&=\max_{\substack{1\leq r\leq p \\ 1\leq j\leq \delta_n\\ t_i\in T_{r}}} \frac{|\hat{H}^{(\ell)}(t_i,s_{r,j},r)|}{\sqrt{\hat{\sigma}^2_{r,t}}}=\max_{\substack{1\leq r\leq p \\ 1\leq j\leq \delta_n\\ t_i\in T_{r}}} \frac{\Bigg|\frac{1}{\sqrt{ns_{r,j}}}\sum_{k =1}^n W\Big(\frac{k/n-t_i}{s_{r,j}}\Big)\Upsilon_{r,k} Z_k^{(\ell)}\Bigg|}{\sqrt{\hat{\sigma}^2_{r,t}}} \label{eqn:gellmax}\\
	\text{crit}_{1-\alpha}(\mathbf{T})&=\text{Quantile}(\{\hat{G}^{(\ell)}_{\text{max}}(\mathbf{T})\}_{\ell=1}^{K_0}\},1-\alpha) \label{eqn:critvalue}
\end{align}
\endgroup
where for each $\ell \in 1,\dots,K_0$ the $\{Z_j^{(\ell)}\}_{j=1}^n$ are iid standard normal random variables. In \eqref{eqn:gellmax}, $\hat{G}^{(\ell)}_{\text{max}}(\mathbf{T})$ is our multiplier bootstrap statistic for the $\ell$'th bootstrap repetition. 
The bandwidth parameter $s'$ controls the size of the block bootstrap, and the selection of an appropriate $s'$ is discussed in \apxref{apx:selecthyper}.  This bootstrap is simple to implement and can be run in parallel to improve the speed of computation. We see in Section \ref{sec:bootstraptheory} that this bootstrap is able to approximate the distribution of \eqref{eqn:mjpdhdmaxstat} asymptotically, enabling the calculation of a critical value at a prescribed level $\alpha \in (0,1)$ in \eqref{eqn:critvalue}. The null hypothesis of no jump is then rejected if $G^{\delta_n}_\text{max}(\mathbf{T}) \geq \text{crit}_{1-\alpha}(\mathbf{T})$. There are multiple reasons why AJDN's simple bootstrap statistic is able to mimic the complicated distribution of its test statistic. First, since the vectors of test and bootstrap statistics $H(t,s,r)$ and $\hat{H}^{(\ell)}(t,s,r)$ are weighted averages, for large $n$ it could be expected their behaviour will be approximately Gaussian, and hence be fully characterized by their mean and covariance. Under the null of no jumps the expected values of $H(t,s,r)$ and $\hat{H}^{(\ell)}(t,s,r)$ will be approximately 0. 
On the other hand, the construction of $\mathbf{\Upsilon}_i$ leads the conditional variance of $\hat{H}^{(\ell)}(t,s,r)$ to be approximately the local long-run variance  of $H(t,s,r)$ in dimension $r$ at time $t$ and the conditional covariance between the test statistics of different dimensions $\hat{H}^{(\ell)}(t,s,r_1),\hat{H}^{(\ell)}(t,s,r_2)$ to be approximately the local long-run cross-covariance of $H(t,s,r)$ between these dimensions, mimicking the covariance structure of the test statistics $H(\cdot)$ across time and dimensions (see \apxref{sec:sprimerot} for a further discussion). 

Detecting multiple jumps involves an iterative algorithm where after detecting a single jump at time $\hat{t}$ in a single dimension $\hat{r}$ with scale $\hat{s}$, AJDN searches for subsequent jumps in all dimensions and times, while excluding a window in $\hat r$-th component of time series around the already detected jump $[\hat{t}-(1+c)\bar{s}_{\hat r},\hat{t}+(1+c)\bar{s}_{\hat r}]$, where we set $c=0.01$. This algorithm continues searching in each dimension, at times not close to previously detected jumps in that dimension, until no more jumps are identified.  An overview of this algorithm is described in Algorithm \ref{alg:mainalgo}. Let $a-1$ denote the number of the detected jumps. Note that for $a=1$, the multiplier bootstrap described in \eqref{eqn:upsilon} to \eqref{eqn:critvalue} is used to calculate $\text{crit}_{1-\alpha}(\mathbf{T}^{(1)})$, and $\hat{G}_{\text{max}}^{(\ell)}(\mathbf{T}^{(1)})$  is calculated over the whole interval and the results $\frac{|\hat{H}^{(\ell)}(t_i,s_{r,j},r)|}{\sqrt{\hat{\sigma}^2_{r,t}}}$  are stored in memory. If $\hat{G}_{\text{max}}^{(\ell)}(\mathbf{T}^{(1)}) > \text{crit}_{1-\alpha}(\mathbf{T}^{(1)})$, we update the detected number of jumps to $1$ by setting $a = 2$. Let the time set $\mathbf{T}^{(2)}$ denote the difference of time set $\mathbf{T}^{(1)}$ and the period in the neighbourhood of the detected change point. Then, $\text{crit}_{1-\alpha}(\mathbf{T}^{(2)})$ can be efficiently calculated by deleting the $\frac{|\hat{H}^{(\ell)}(t_i,s_{r,j},r)|}{\sqrt{\hat{\sigma}^2_{r,t}}}$ statistics in a neighbourhood around the previously detected jump, and utilizing the remaining statistics. If $\hat{G}_{\text{max}}^{(\ell)}(\mathbf{T}^{(2)}) > \text{crit}_{1-\alpha}(\mathbf{T}^{(2)})$, we increase $a$ by one and continued the aforementioned procedure, until the algorithm ends. This algorithm yields  a list (which is empty in the case of no jumps) $J$, where for the $j$-th jump the list includes the estimated time of change $\hat{t}_j$, the dimension $\hat{r}_j$, and the scale that the jump was detected with $\hat{s}_j$.

\begin{remark}
We see in the theoretical results of Section \ref{sec:asympresults} that our proposed test, which uses a specially designed sparse sequence of scales to approximate the statistic in \eqref{eqn:firstmaxstatistic}, can consistently detect multiple jumps with a controlled Type I error rate under both smoothly and abruptly time-varying temporal dynamics. Our results are built on new high-dimensional Gaussian approximation results on hyperrectangles. Compared with the seminal work in \cite{chernozhukov2017central}, our results further allow complex temporal dependence between the high-dimensional vectors, as well as different sparsities in different dimensions of these vectors, which is the key to mimicking the asymptotic behavior of the multiscale nonparametric statistics to achieve a consistent test of jumps with asymptotic correct size, admitting diverging number of jump points. The new Gaussian approximation theory is of separate interest and can serve as a general tool for the analysis of high-dimensional and nonstationary time series. 
\end{remark}
\begin{algorithm}[t]
\SetAlgoLined
 Output: $J$, the set of jump points\; 
 Initialization: 
 $J\leftarrow \emptyset$,
 $a=1$,
 $\mathbf{T}^{(a)}=\mathbf{T}$\;
 \While{$G_{\emph{max}}^{\delta_n}(\mathbf{T}^{(a)})\geq \emph{crit}_{1-\alpha}(\mathbf{T}^{(a)})$}{
  $\{\hat{t},\hat{s},\hat{r}\} \leftarrow \text{argmax}_{1\leq r\leq p,\;1\leq j\leq \delta_n,\;t_i\in T_r} G(t_i,s_{r,j},r)$\;
  $J=J\cup \{\hat{t},\hat{s},\hat{r}\}$\;
  \For{$r\neq \hat{r}$}{
	$T_r^{(a+1)}=T_r^{(a)}$  
  }
  $T_{\hat{r}}^{(a+1)}=T_{\hat{r}}^{(a)} \setminus [\hat{t}-(1+c)\bar{s}_r,\hat{t}+(1+c)\bar{s}_r]$\;
  $a=a+1$
 }
 \caption{AJDN-H: First Stage AJDN}\label{alg:mainalgo}
\end{algorithm}




\subsection{Second stage localization of AJDN.}\ 
Once jumps have been detected a second stage procedure is implemented to improve the estimation of the locations $\hat{t}_j$ for all $j$. Second stage refinement in jump detection has been shown to be optimal in the one dimensional case. For instance, \cite{muller1992change} and \cite{gao2008nonparametric} derived the optimality of the two-stage method assuming the number of jumps is known and is bounded. 
\cite{wu2019multiscale} further generalize  \cite{muller1992change} and  \cite{gao2008nonparametric}  under complex temporal dynamics, showing that the second stage refinement is optimal when the number of jumps is unknown and diverging.

For high-dimensional time series in this paper, the second stage refinement utilizes local CUSUM statistics around the estimated jumps. For each jump $j$ in dimension $r$ with estimated location $\hat{d}_{r,j}$ let
\begin{align}
	S_{r,I}&=\sum_{i\in \lambda(I)}y_{r,i}, \;\;\; \lambda(I)=\{i\;|\; i/n \in I\},\;\;\;|\lambda(I)|=\text{Cardinality of }\lambda(I)
\end{align}
and for some $z_n>0$, and $\tilde{\alpha}>-1$ (we set $\tilde{\alpha}=-0.5$ and $z_n=\underline{s}_\text{min}/2$, where $\underline s_{\min}=\min_{1\leq r\leq p}\underline s_r$), 
\begin{align}
	l_{r,j}=\hat{d}_{r,j}-(2+\tilde{\alpha})z_n,\;\;u_{r,j}=\hat{d}_{r,j}+(2+\tilde{\alpha})z_n,\;\;\tilde{l}_{r,j}=\hat{d}_{r,j}-z_n,\;\;\tilde{u}_{r,j}=\hat{d}_{r,j}+z_n.
\end{align}
Then the local CUSUM statistic $V_j(t)$, and the refined jump estimate $\tilde{d}_{r,v}$ is 
\begin{align}
V_{r,j}(t)=S_{r,[l_{r,j},t]}-\frac{|\lambda([l_{r,j},t])|}{|\lambda([l_{r,j},u_{r,j}])|}S_{r,[l_{r,j},u_{r,j}]}\;\;\;\tilde{d}_{r,j}=\text{argmax}_{t\in[\tilde{l}_{r,j},\tilde{u}_{r,j}]}|V_{r,j}(t)|.
\end{align}

Here the CUSUM statistics are calculated using the wide window $[l_{r,j},u_{r,j}]$ around $\hat{d}_{r,j}$, and then the maximum is taken within a narrower window $[\tilde{l}_{r,j},\tilde{u}_{r,j}]$ around $\hat{d}_{r,j}$. After the second stage refinement step we see in Section \ref{sec:highdimlocalcusum} that asymptotically the Type I error rate of AJDN is still controlled.
\subsection{Estimating  mean functions via homogeneity pursuit.}\ 
Let $b$ be a bandwidth parameter for preliminary estimation, $h$ be the bandwidth parameter for the final nonparametric smoothing. In particular, $b/h \to 0$. We propose the following procedures.\\
\noindent\textbf{(a) Computing time stamps and corresponding intervals.}
\begin{description}
\item(i) Obtain estimated time stamps of jumps of each dimension $1 \leq r \leq p$, namely $\hat d_{r, 1}< \cdots< \hat d_{r, \hat m_{r,n}}$. Set $\hat d_{r, 0} = 0$, $\hat d_{r, \hat m_{r,n}+1} = 1$. 
\item(ii) Combine jumps $\hat d_{r_1, v_1}$ and $\hat d_{r_2, v_2}$, if $|\hat d_{r_1, v_1} - \hat d_{r_2, v_2}|  <  s_n$, and obtain a sequence of unique ascending jumps of all dimensions, $0 = \hat d_{0} < \hat d_{1}< \cdots< \hat d_{\hat m_n}< \hat d_{\hat m_n + 1}=1$. 
 \end{description}

In the above, (ii) combines nearby jump points, yielding a pasimonus high-dimensional piecewise smooth trends structure, allowing the possible asynchronous jumps. To present the next step, let $\hat{\mathcal I}_{r, q} = (\hat d_{r,q}, \hat d_{r, q+1}]$, $q=0,\cdots, \hat m_{r,n}  $, $\hat{\mathcal I}_{l} = (\hat d_l, \hat d_{l+1}]$, $\hat{\mathcal I}^b_{l} =  [\hat d_l + b, \hat d_{l+1} - b] \cap [0,1]$, $l=0,\cdots, \hat m_{n}$. 

\noindent \textbf{(b) Estimating spatial group structures of trend}. 
We consider the hierarchical clustering method, see \cite{murtagh2012algorithms} for a thorough review.  We employ the kernel-based distance. First, for $j=1, \cdots, p$, we obtain the kernel estimator 
 \begin{align}
     (\check \beta_j(t,b),\check \beta^{\prime}_{j}(t,b)) ^{\top}=    {\arg\min}_{\eta_0, \eta_1 \in  \R}  \sum_{i/n \in \hat{\mathcal I}_{l}^b}[Y_{i,j} - \eta_0 -  (t-i/n)\eta_1(t)]^2 K_b(i/n,t),    \notag\end{align}
where $K(\cdot)$ is a kernel function (see also 
Assumption \ref{ass:kernel}, typically one can choose Epanechnikov kernel), and $K_b(\cdot) = K(\cdot/b)$. 
 Then, we combine the two estimators using bandwidth $b/\sqrt{2}$ and $b$ to reduce the bias
 \begin{align}
     \tilde \beta_j(t) = 2 \check \beta_j(t,b/\sqrt{2}) - \check \beta_j(t,b). \notag\label{eq:jackbeta}
 \end{align}
Then, the kernel-based distance is defined as
\begin{align}
    \tilde \Delta_{ij, l} = \frac{1}{n}\sum_{ k=1}^n|\tilde \beta_{i}(k/n) - \tilde \beta_{j}(k/n)| \mf 1(k/n \in \hat{\mathcal I}^b_{l}), \notag
\end{align}
and the distance matrix $\tilde{\bs  \Delta}_{n,l} = ( \tilde \Delta_{ij, l})_{i,j=1}^n$. We shall show that $\tilde \Delta_{ij, l}$ is an estimate of 
\begin{align}
   \Delta_{ij, l}^{\circ} = \int_{t \in \mathcal I_l} |\beta_{i}(t) - \beta_{j}(t)| dt ,
\end{align}
\begin{enumerate}
    \item On each $\hat{\mathcal I}_{l}$. Start with $p$ clusters
    \item  For two difference clusters $A$ and $B$, the distance between $A$ and $B$ is $\max_{i \in A, j \in B}  \tilde \Delta_{ij, l}$. Merge the two clusters with the smallest distance. Recalculate the distance between different clusters and update the distance matrix. 
    \item Repeat Step 1 and 2 until the number of clusters reaches $K_l^0$ and obtain $\tilde C_{1,l},\ldots, \tilde C_{K_l^0,l}$, which yields a tree of nested groups and can be drawn into a dendrogram.
    \item Let $n_l = \hat d_{l+1}- \hat d_l$. The number of the groups is calculated by minimizing information criterion 
    \begin{align}
        \mathrm{IC}(K, l) = \log (\tilde \sigma_{l}^2(K) )+ K \chi_n,
    \end{align}
    where $\chi_n \to 0$ is a tuning parameter for penalizing the group number, 
    \begin{align}
        \tilde \sigma_{l}^2(K)  = \frac{1}{n_l b p}\sum_{k = 1}^K \sum_{j \in \hat{C}_{k,l}}\sum_{i/n \in \hat{\mathcal I}_{l}^b}  [Y_{i,j} - \tilde \alpha_{k,l}(i/n)]^2 ,\label{eq:sigmal}
    \end{align}
    where $\tilde \alpha_{k,l}(t)$ is obtained from local linear estimation and Jackknife correction. Specifically, we first obtained the estimate from the local linear estimation with bandwidth $b/\sqrt{2}$ and $b$ using
    \begin{align}
     (\check \alpha_{k,l}(t,b),\check \alpha^{\prime}_{k,l}(t,b)) ^{\top}=    {\arg\min}_{\eta_0, \eta_1 \in  \R}  \sum_{i/n \in \hat{\mathcal I}_{l}^b}[\bar Y_{i,k} - \eta_0 -  (t-i/n)\eta_1(t)]^2 K_b(i/n,t),    \label{def:alpha}
     \end{align}
where $\bar Y_{i,k} = \sum_{j \in \hat{C}_{k,l}}
 Y_{i,j}/|\hat{C}_{k,l}|$. Then, we combine the two estimators to reduce the bias
 \begin{align}
     \tilde \alpha_{k,l}(t) = 2 \check \alpha_{k,l}(t,b/\sqrt{2}) - \check \alpha_{k,l}(t,b). \label{eq:jackalpha}
 \end{align}
    \item Let $\tilde K_l =  {\arg\min}_{1 \leq k \leq \bar K}  \mathrm{IC}(k,l)$, for a prespecified number of groups $\bar K$. Output the estimated groups $\hat C_{1,l}, \ldots, \hat C_{\tilde K_l,l}$. 

\end{enumerate}
   \textbf{Homogeneity pursuit in estimation.} 
   Estimate the mean function for each dimension based on the estimated group structures. For $1 \leq q \leq \hat m_{r}+1$, $1 \leq r \leq p$, 
   \begin{align}
       (\hat \beta_{r,q} (t), \hat \beta^{\prime}_{r,q}(t))^{\top} &= \arg \min_{\eta_0, \eta_1 \in \mathbb R} \sum_{i=1}^n \sum_{l=1}^{\hat m_n+1}\mf 1(i/n \in \hat{\mathcal  I}_l \cap \hat{\mathcal  I}_{r,q}) \notag \\ &\times \left[\sum_{k=1}^{\tilde K_l}  \{\sum_{j \in \hat{C}_{k,l}}
 Y_{i,j}/|\hat{C}_{k,l}| - \eta_0 -  (t-i/n)\eta_1(t)\}^2 K_h(i/n,t)  \mf 1(r \in \hat{C}_{k,l}).  \right]\label{def:hatbeta}
   \end{align}
   The final estimator is 
   \begin{align}
       \hat \beta_{r} (t) =\sum_{q=1}^{\hat m_{r,n}+1}\hat \beta_{r,q} (t) \mf 1(t \in \hat{\mathcal I}_{r,q}).
   \end{align}
   \begin{remark}
The algorithm can be extended for estimating paralleled mean functions via homogeneity pursuit. We can replace the $\tilde \beta_j(t)$ with $\tilde \beta_j^{-}(t) =\tilde \beta_j(t) - \int_{t \in \hat{\mathcal I}_l^b} \tilde \beta_j(t)dt $, for $t \in \hat{\mathcal I}_l^b$, on each $\mathcal I_l$ replace $\beta_j(t)$ with $\beta_j^{-}(t) = \beta_j(t) - \int_{\mathcal I_l} \beta_j(t) dt$, $Y_{i,l}$ in \eqref{eq:sigmal}, \eqref{def:alpha} and \eqref{def:hatbeta} with $Y_{i,j} - \int_{t \in \hat{\mathcal I}_l^b} \tilde \beta_j(t)dt $. Finally, the estimator is $ \hat \beta_{r} (t) =\sum_{q=1}^{\hat m_{r,n}+1}\hat \beta_{r,q} (t) \mf 1(t \in \hat{\mathcal I}_{r,q}) + \sum_{l=1}^{\hat m_n+1} \mf 1(t \in \hat{\mathcal I}_l^b)\int_{\hat{\mathcal I}_l^b} \tilde \beta_r(s) ds $
   \end{remark}

\section{Assumptions for AJDN.}\label{sec:assumptions}

\subsection{Assumptions on mean functions.}\label{sec:assumptionjumps}\ 
We assume that the $p$-dimensional mean function $\bs \beta(t)=(\beta_1(t),...,\beta_p(t))^\top$ satisfy the following conditions. For $1\leq r\leq p$, there exist constants $k$, $C_{Lip}$, $\bar C$ and $p$ sequences $\gamma_{r,n}$ and $\Delta_{r,n}$ such that
\begin{description}
	\item (M1) 
	$\beta_r(\cdot)$ has $m_{r,n}$ discontinuous points $0=d_{r,0}<d_{r,1}<d_{r,2}<...<d_{r,m_{r,n}}<d_{r, m_{r,n}+1}=1$, and $\beta_r(x)\in C^k((d_{r,i},d_{r,i+1}),C_{Lip})$ for $x\in (d_{r,i},d_{r,i+1}) $, $0\leq i\leq m_{r,n}$. In addition, $\beta_r(\cdot)$ is either right or left continuous at $\{d_{r,i},0\leq i\leq m_{r,n}+1\}$. Also $m_{r,n}=0$ corresponds to no break points in $\beta_r(\cdot)$.
	\item (M2) 
	$\Delta_{r,i,n}:=|\beta_r(d_{r,i}+)-\beta_r(d_{r,i}-)|\geq \Delta_{r,n}>0$ for $1\leq i\leq m_{r,n}$, where for any function $g$ and $a\in\mathbb R$, $g(a+)=\lim_{s\downarrow a} g(s)$ and $g(a-)=\lim_{s\uparrow a} g(s)$. 
	\item (M3) 
	$\min_{0\leq i\leq m_{r,n}} |d_{r,i+1}-d_{r,i}|\geq \gamma_{r,n}>0$ for all $r$  such that $\beta_r$ is piecewise smooth.
	\item (M4) $\sup_{a\in [0,1)}|
	\beta_r^{(u)}(a+)|\leq \bar C,\sup_{a\in (0,1]}|
	\beta_r^{(u)}(a-)|\leq \bar C, \  \forall~ 0\leq u\leq k$.
\end{description}
Condition (M1) means that each component of $\bs \beta(\cdot)$  is piece-wise smooth, with Lipschitz continuous $k_{th}$ order derivative. 
(M2) and (M3) assume a lower bound  $\Delta_{r,n}$ and $\gamma_{r,n}$ on the minimum jump size and minimum spacing between two jump points in the same dimension. 
Condition (M4) controls the overall smoothness of derivatives of $\beta_r(\cdot)$ for $1\leq r \leq p$,  which matters when the number of jumps, $m_{r,n}$, diverges. 

\subsection{Assumptions on high-dimensional non-stationary errors.\ }\label{sec:apls}
\ Assume that the error process $\bs \epsilon_{i,n}$ is a $p$-dimensional piecewise locally stationary (PLS) process defined in \cite{zhou2013heteroscedasticity} which can experience both abrupt and smooth changes in its data generating mechanism over time. Specifically, $\bs \epsilon_{i,n}=\mf L(t_i,\FF_i)=(L_r(t_i,\FF_{i}))_{1\leq r\leq p}$ where  $\FF_i=(\eta_{-\infty},...,\eta_i)$ is the filtration generated by $i.i.d$ random elements $\eta_i$ defined on some measurable space $\mathcal S$,   $\mf L: [0,1]\times \mathcal S^\mathbb Z \rightarrow \mathbb R^p $ is a measurable function,  and $\epsilon_{r,i}$ and $L_r$ are the $r_{th}$ components of $\bs \epsilon_i$ and $\mf L$, respectively, such that $\epsilon_{r,i}=L_r(t_i,\FF_i)$. Let $\mathcal P_j=\E(\cdot|\FF_j)-\E(\cdot|\FF_{j-1})$ be the projection operator. For any $\FF_i$ measurable random variables $g=g(\FF_i)$, write $g^{(j)}=g(\FF_i^{(j)})$ where $\FF_i^{(j)}$ is the filtration obtained by replacing the $j_{th}$ innovation of $\FF_i$ via its $i.i.d.$ copy.
 
 \begin{description}  
 	\item(A1)(Piece-wise locally stationary components) $(\epsilon_{i,r})_{i=1}^n$ has  $l_r$ $(l_r\geq 0)$ break points $\{c_{r,s}\}_{1\leq s\leq l_r}$ satisfying  $0=c_{r,0}<c_{r,1}<...<c_{r,l_r}<c_{r,l_r+1}=1$  such that 
 	$L_{r}(t,\FF_i)=L_{r,j}(t,\FF_i)$ if $c_{r,j}<t\leq c_{r,j+1}$, where $\{L_{r,s}, 0\leq s\leq l_r\}$  are measurable functions.  In particular,
 		\begin{align}
 		\epsilon_{r,i}=L_r(i/n,\FF_i)=L_{r,j}(i/n,\FF_i),\  c_{r,j}<i/n\leq c_{r,j+1}, 0\leq j\leq l_r,
 		\end{align} where uniformly for $1\leq r\leq p$,
 		\begin{align}\label{orignial-A3}
 		\|L_{r,j}(t,\FF_0)-L_{r,j}(s,\FF_0)\|_2\leq C|t-s|
 		\end{align}
 		for all $t,s\in (c_{r,j},c_{r,j+1}]$, $0\leq j\leq l_r$, and for some universal finite constant $C$.
 	\item(A2) $\E(L_r(t,\FF_0))=0$ for $1\leq r\leq p$ and $t\in [0,1]$. Furthermore, there exists a positive constant $t_0>0$ such that  $\max_{1\leq r\leq p}\sup_{t\in (0,1]}\E(\exp(t_0 |L_{r}(t,\FF_0)|))<\infty.$
 	\item(A3) Define  the dependence measure of $L$  in $\mathcal L^\kappa$ as
 	\begin{align}\delta_{\kappa}(L,i):=\max_{1\leq r\leq p}\sup_{t\in (0,1]}\|L_r(t,\FF_i)-L_r(t,\FF_i^{(0)})\|_{\kappa}.\end{align} 
  Assume that
 	 $\delta_1(L,i)=O(\chi^i)$ for some $\chi\in (0,1)$.
 	
 	\item(A4) For $1\leq r\leq p$, the long-run variance $\sigma_{lrv, r}^2(t)$ of $(\varepsilon_{i,r})_{1\leq i\leq n}$ is Lipschitz continuous on $(c_{r,j},c_{r, j+1}]$ for $0\leq j\leq l_r$, $\max_{1\leq r\leq p}\sup_{t\in[0,1]}\sigma_{lrv, r}^2(t)\leq M<\infty$ for some constant $M$, and $\min_{1\leq r\leq p}\inf_{t\in[0,1]}\sigma_{lrv, r}^2(t)>0,$ where $\sigma_{lrv,r}^2(0)=\lim_{t\downarrow 0}\sigma_{lrv,r}^2(t)$ and \begin{align}\label{longrun}
 	\sigma^2_{lrv,r}(t):=\sum_{k\in \mathbb Z}Cov(L_r(t,\FF_0),L_r(t,\FF_k)), 0<t\leq 1.
 	\end{align}
 	\item(A5) For $1\leq r\leq p$, the  derivative of variance $\sigma_{ r}^2(t)=Var(L_r(t,\FF_0))$ is Lipschitz continuous on $(c_{r,j},c_{r, j+1}]$ for $0\leq j\leq l_r$. Besides, there exist constants $0<\underline \sigma\leq \bar \sigma<\infty$ such that $\max_{1\leq r\leq p}\sigma^2_{r}(t)\leq \bar \sigma^2<\infty,$  $\min_{1\leq r\leq p}\inf_{t\in[0,1]}\sigma^2_{r}(t)\geq \underline \sigma^2>0,$ where $\sigma_{r}^2(0)=\lim_{t\downarrow 0}\sigma_{r}^2(t)$. 
 \end{description} 

Condition (A1) defines the high-dimensional piecewise locally stationary (PLS) process, and the filters $L_r$, $1\leq r\leq p$  are stochastic piecewise lipschitz continous.   (A2) assumes the errors have sub-exponential tails. For condition (A3), the quantity $\delta_{\kappa}(L,i)$  is called ``physical dependence measures'' quantifying the dependence of $\mf L(t,\FF_i)$ on $\eta_0$ and is assumed to decay geometrically to zero.  For presentational simplicity, we shall demonstrate all our results under the geometrical decay assumption.  We can establish the theoretical results of the paper when $\delta_{\kappa}(L,i)$ decays at a sufficiently fast polynomial rate with substantially more involved mathematical arguments.  We refer to \cite{wu2018gradient} regarding the calculations of $\delta_{\kappa}(L,i)$ for many PLS linear and nonlinear processes. 
 Conditions (A4) and (A5) guarantee that the long-run variance and variance of the time series in each component are non-degenerate over $[0,1]$, and are piece-wise Lipschitz continuous.
 

\subsection{Assumptions on nonlinear filters.}\ \label{sec:nonlinearfilter}
We adopt the following filter class defined in \cite{wu2019multiscale}. For each positive integer $k$, 
let $\mathcal W(k)$ be the collection of functions $W$ satisfying
\begin{align}
&W\in C^{1} (\mathbb R, C_{Lip}),\  \text{\it Supp}(W)\subseteq[-1,1],\ W(x)=-W(-x), \int_{0}^1 W(x)dx=1,  \notag\\&\lim_{x\downarrow -1} W'(x)=\lim_{x\uparrow 1} W'(x)=0,  \int_{-1}^1 x^uW(x)dx=0 \ \ \mbox{for}\ \  1\leq u\leq k
\end{align}
where $C_{Lip}>0$ is some constant. 
 The filter $W(\cdot)$ is the $k_{th}$  order jump-pass filter  ($k\geq 2$) if it satisfies 
\begin{description}
	\item (W1) $W(\cdot)\in \mathcal W(k)$.
	\item (W2) Define $F_w(x)=\int_{-1}^xW(s)ds$ such that i)  $0$ is the unique maximizer of $|F_w(x)|$, and $|F_w(0)|$ exceeds all other local maximum by some  positive constant $\bar \eta_0$;  ii) $F^2_w(t)-F^2_w(0)\leq -\bar{\eta}_1 t^2$ for $|t|\leq \bar{\eta}_2$  for some strictly positive constants $\bar{\eta}_1$ and $\bar{\eta}_2$,  and  $W'(0)\neq 0$.
\end{description}
\cite{wu2019multiscale} proves that there are always piecewise polynomial functions satisfying (W1) and (W2) for any order $k\geq 2$. Condition (W1) guarantees that $W(\cdot)$ filters out smooth signals from the data, leading to negligible remaining terms. Filters with higher order $k$ will yield a smaller order of the remaining signal, but also often result in larger noise. We refer to \cite{wu2019multiscale}  for more details, where they propose schemes for choosing polynomial filters according to signal noise ratios. 

\subsection{Assumptions on scales.\ }\label{sec:assumptiononscale}
Recall $\bar s_{\max}=\max_r \bar s_r$ and $\bar s_{\min}=\min_r \bar s_r$ and consider that $\bar s_{\min }\asymp n^{-\iota_0}$, $\bar s_{\max }\asymp n^{-\iota_1}$
for some $0<\iota_1\leq \iota_0<1$. Let $T_{r,c}=\cup_{0\leq i\leq l_r} (c_{r,i}+\bar s_r, c_{r,{i+1}}-\bar s_r)$, $\bar T_{r,c}=\cap_{1\leq i\leq l_r} [c_{r,i}-\bar s_r, c_{r,i}+\bar s_r] $, $T_{r,d}=\cup_{0\leq i\leq m_{r,n}} (d_{r,i}+\bar s_r, d_{r,{i+1}}-\bar s_r)$, $\bar T_{r,d}=\cap_{1\leq i\leq m_{r,n}} [d_{r,i}-\bar s_r, d_{r,i}+\bar s_r] $. 
We have the following assumptions for the scales $\bar s_r$ and $\underline s_r$, $1\le r\le p$ and for $s'$.
\begin{description}
	\item (B1) If $m_{r,n}\geq 1$, then  $\bar {s}_r\leq d_{r,1}\leq d_{r,m_{r,n}}\leq 1-\bar {s}_r$. If further $m_{r,n}\geq 2$, then additionally $\bar {s}_r\leq \min_{2\leq i\leq m_{r,n}} |d_{r,i}-d_{r,i-1}|/2$.  For $s'$, assume $s'=o(1)$, $ns'\rightarrow \infty$, $s'\max_r l_r/\min_r \underline s_r=o(1)$. 
	\item (B2) $m_{r,n}\bar {s}_r=o(1)$.
	\item (B3) $\sqrt{n\bar {s}_r}\bar {s}_r^{k+1}=o(1)$ and $\frac{m_{r,n}}{n\underline {s}_r}=o(1)$.
\end{description}
 Assumption (B1) requires that $2\bar {s}_r$  are smaller than the minimal spacing of jump points in the same dimension. Condition (B2) 
means that the lengths of the intervals  $\bar T_r^d$ are asymptotically negligible. The break points $(c_{r,i})$ and  $(d_{r,i})$ are allowed to be overlapped. For (B3), the term $\sqrt{n\bar {s}_r}\bar {s}_r^{k+1}$ is the bias caused by the $k_{th}$ order jump-pass filter, while the term $\frac{m_{r,n}}{n\underline {s}_r}$ controls the approximation errors of the Riemann sums due to the observations at discrete times. 
The best scales to capture the jumps at different time points of the $r_{th}$ component of the $p$-dimensional non-stationary time series are usually different but will fall within $[\underline s_r, \bar s_r]$ if the intervals are sufficiently wide.

\section{Asymptotic results.}\label{sec:asympresults}
\ We first present some notation. For two positive real sequences $a_n$ and $b_n$, write $a_n\asymp b_n$ if there exist constants $M_1$ and $M_2$, $0<M_1\leq M_2<\infty$, such that $M_1\leq \liminf a_n/b_n\leq \limsup a_n/b_n\leq M_2$, and $a_n\lessapprox b_n$ if there exists a universal constant $M$ such that $a_n\leq M b_n$. Let $C$ be a generic constant which varies from line to line and does not rely on time and dimension. Write $t_i=i/n$. For any vector $\mf Z=(z_1,...,z_p)^\top$ let $|\mf Z|_\infty=\max_{1\leq i\leq p}|z_i|$, $|\mf Z|=(\sum_{i=1}^p z_i^2)^{\frac{1}{2}}$ be its Euclidean norm, and $|\mf Z|_e=(|z_1|,...,|z_p|)^\top$.  For a random vector $\mathbf v$, denoted by $\|\mathbf v\|_q=\left(\E |\mathbf v|^q\right)^{\frac{1}{q}}$ for its $\mathcal L^q$ norm, and we omit the subscript $q$ when $q=2$. In the theoretical results, without loss of generality, we assume that $p\geq 9$ so $\log p>2$, which will simplify the argument for using Burkholder inequality. For any $p$-dimensional vector $\mf v=(v_1,...,v_{p})^\top$ and any index set $A\in \{1,...,p\}$, let $\mf v[A]$ be the vector by deleting $v_i$ from $\mf v$ if $i\not \in A$. For vectors $\mf v_1,..\mf v_k$, let $vec(\mf v_1,...,\mf v_k)$ be the long vector $(\mf v_1^\top,...,\mf v_k^\top)^\top$ which stacks $\mf v_i's$. Write $C^k(I, C_{Lip})$ for the collection of continuous functions with $k_{th}$ Lipschitz continuous derivatives on interval $I$ with respect to the Lipschitz constant $C_{Lip}$.
When indices discussed involve the jump points, by default we only consider those dimensions when jump points exist if no confusion is caused. For example, the notation $\max_{r,v}|\hat d_{r,v}-d_{r,v}|$ means the maximum deviation of the estimated locations of jump points, and the range of $r$ in the maximum excludes $k$ if $\beta_k(\cdot)$ is smooth. Let $\mf 1(\cdot)$ be the indicator function.

\subsection{Sparsification and discretization.}\ \label{sec:sd} Observe that 
\begin{align}\label{decomposition}
G(t,s,r)=|H(t,s,r)|/\sqrt{\hat \sigma^2_{r,t}}=|\tilde H(t,s,r)+\tilde G(t,s,r)|/\sqrt{\hat \sigma^2_{r,t}},
\end{align}
where  \begin{align}
\tilde H(t,s,r)&=\frac{1}{\sqrt{ns}}\sum_{j=1}^n\epsilon_{r,j} W\left(\frac{j/n-t}{s}\right),~
\tilde G(t,s,r)=\frac{1}{\sqrt{ns}}\sum_{j=1}^n\beta_{r}(j/n) W\left(\frac{j/n-t}{s}\right).\notag
\end{align}
In \apxref{Uniform-Consistency-sigma} we show that $\hat \sigma_{r,t}^2$ converging uniformly to $\E \hat \sigma_{r,t}^2$ on $T_r$. 
Moreover, $\E \hat \sigma_{r,t}^2$ is always bounded below and above by positive constants. These fact helps us to evaluate $\frac{\tilde H(t,s,r)}{\sqrt{\hat \sigma^2_{r,t}}}$. 
\begin{theorem}\label{discrete} 
Under conditions (M) with $k\geq 2$, (A) and (B) in Section \ref{sec:assumptions}, and that $W(\cdot)\in \mathcal W(k)$ for some $k\geq 2$, we have that  there exists a constant $C$ such that for any $x>0$,

\begin{align}
&\p\left(\left|\sup_{\substack{1\leq r\leq p \\ \underline{s}_r\leq s\leq \bar{s}_r\\ t\in I_r, I_r\in  T_r, }}\frac{|\tilde  H(t,s,r)|}{\sqrt{\hat{\sigma}_{r,t}^2}}-\sup_{\substack{1\leq r\leq p \\ 1\leq j\leq \delta_n\\ t_i\in I_r, I_r \in T_r}} \frac{|\tilde H(t_i,s_{r,j},r)|}{\sqrt{\hat{\sigma}_{r,t_i}^2}}\right|\geq x\right) \notag\\
&\leq\left(\frac{C[\log^{5/2}(pn)+(\max_r\log (\bar s_r/\underline s_r)\vee 1)\log^{9/2}(pn)]}{xn\underline s_{min}}+\frac{C}{x\log^{\epsilon} n}+v_n^{-1}\right)^{\log (pn)},
\end{align}
where $I_r$ is any sub-interval of $T_r$, $\epsilon>1/2$ is defined in \eqref{sparsescaledelta}, and $v_n$ is a sequence satisfying that $v_n\rightarrow \infty$ as well as that $v_n(n\bar s_{\min})^{-1/2}\log^{7/2} (pn)=o(1)$. If $p=O( n^\iota)  $ for some $\iota>0$ then the upper bound can be simplified to  $(C/(x\log^\epsilon n)+v_n^{-1})^{\log(pn)}$.
\end{theorem}
Theorem \ref{discrete} shows  that we can compute the maximum test statistic at discrete times and for the sparse sequence of scales \eqref{sparsescale}. The discretized maximum test statistic will be close to $G_{\text{max}}(\mathbf{T})$ for high-dimensional time series when the dimension $p\asymp \exp(n^\iota)$ for some $\iota>0$.  Moreover, the discretization enables us to utilize the Gaussian approximation results developed in \apxref{Gaussian-approximation} for sparse high-dimensional nonstationary time series. Compared to the state-of-the-art Gaussian approximation results in \cite{dette2021confidence}, the new Gaussian approximation results in this paper further admit dramatically different sparsities among different dimensions and are valid on hyperrectangles,  both of which are essential for the multiscale inference in this paper.

\subsection{Limiting behavior of the test statistics.}\ 
Recall $\tilde H(t,s,r)=\frac{1}{\sqrt{ns}}\sum_{j=1}^n\epsilon_{r,j} W\left(\frac{j/n-t}{s}\right)$. Let $\breve \by_1$,...,$\breve \by_n$ be a
 a sequence of centred 
$p$-dimensional  Gaussian vectors with the same auto-covariance structure as  the vectors $(\bepsilon_i)_{1\leq i\leq n}$, and $\breve \by_i=(\breve y_{r,i})_{1\leq r\leq p}$. Let $\tilde H^y(t,s,r)=\frac{1}{\sqrt{ns}}\sum_{j=1}^n\breve y_{r,j} W\left(\frac{j/n-t}{s}\right)$. 
\begin{theorem}\label{Thm1}
Assume condition (A) in Section \ref{sec:assumptions} for $p$-dimensional time series $(\bepsilon_{i})_{1\leq i\leq n}$ where $p=O(n^\iota)$ for some $\iota>0$, (M) with $k\geq 2$, (B), and that $W(\cdot)\in \mathcal W(k)$ for some $k\geq 2$. Assume that $\delta_n=O(n^\iota)$ for some $\iota>0$, and that there exist constant $\iota_1>\iota_0$ such that $d_1n^{-\iota_1}\leq \underline s_{\min }\leq  \bar s_{\max} <d_0n^{-\iota_0}$ for some small positive constant $d_1$ and large constant $d_0$.  
\begin{align}
	&\sup_{\substack{I_r\in T_{r,d},1\leq r\leq p, \\x\in \mathbb R}}\big|\p\big(\sup_{\substack{1\leq r\leq p \\ \underline{s}_r\leq s\leq \bar{s}_r\\ t\in I_r}} \frac{|H(t,s,r)|}{\sqrt{\hat{\sigma}_{r,t}^2}}\leq x\big)-\p\big(\sup_{\substack{1\leq r\leq p \\ 1\leq j\leq \delta_n\\ t_i\in I_r}} \frac{|\tilde H^y(t_i,s_{r,j},r)|}{ \sqrt{\E\hat{\sigma}_{r,t_i}^2}}\leq x\big)\big|\notag\\&=O(\log^{-(\epsilon-1/2)\eta} n) \notag 
\end{align}
for any $0<\eta<1$, where $\epsilon>1/2$ is defined in \eqref{sparsescaledelta}.
\end{theorem}
Theorem \ref{Thm1} shows that the distribution of $\sup_{\substack{1\leq r\leq p \\ \underline{s}_r\leq s\leq \bar{s}_r\\ t\in I_r}} \frac{|H(t,s,r)|}{\sqrt{\hat{\sigma}_{r,t}^2}}$ is well approximated by $\sup_{\substack{1\leq r\leq p \\ 1\leq j\leq \delta_n\\ t_i\in I_r}} \frac{|\tilde H^y(t_i,s_{r,j},r)|}{ \sqrt{\E\hat{\sigma}_{r,t_i}^2}}$, the maximum of  $O(np\delta_n)$ absolute values of Gaussian random variables, or equivalently the maximum norm of a high-dimensional Gaussian vector.  Note that Theorem \ref{discrete} indicates that $\sup_{\substack{1\leq r\leq p \\ \underline{s}_r\leq s\leq \bar{s}_r\\ t\in I_r}} \frac{|H(t,s,r)|}{\sqrt{\hat{\sigma}_{r,t}^2}}$ can be approximated by $\sup_{\substack{1\leq r\leq p \\ 1\leq j\leq \delta_n\\ t_i\in I_r, I_r \in T_r}} \frac{|\tilde H(t_i,s_{r,j},r)|}{\sqrt{\hat \sigma_{r,t_i}^2}}$, which can be further written as a weighted average of high-dimensional time series since every $\hat \sigma_{r,t_i}^2$ converges to a deterministic number and these numbers are bounded from below and above. We refer to \apxref{Uniform-Consistency-sigma} for more details. Using this fact, Theorem \ref{Thm1} can be proved via the Gaussian approximation results of \apxref{Gaussian-approximation}.  We stress that we control the difference of the distribution on all possible $I_r\in T_{r,d}$, which makes our results more general than approximating $G_{\max}(\mf T)$ in \eqref{eqn:firstmaxstatistic} when no jumps are present. This is necessary for the further investigation of Algorithm \ref{alg:mainalgo}. Notice that the rate $\log^{-(\epsilon-1/2)\eta}n$ can be improved if denser scales are used at a more expensive computational cost. On the other hand, the limiting distribution of $\sup_{\substack{1\leq r\leq p \\ 1\leq j\leq \delta_n\\ t_i\in I_r}} \frac{|\tilde H^y(t_i,s,r)|}{ \sqrt{\E\hat{\sigma}_{r,t_i}^2}}$ is still difficult to evaluate. Therefore in the next Section, we shall evaluate this distribution via the bootstrap method which mimics the stochastic behavior of  $\sup_{\substack{1\leq r\leq p \\ 1\leq j\leq \delta_n\\ t_i\in I_r}} \frac{|\tilde H^y(t_i,s,r)|}{ \sqrt{\E\hat{\sigma}_{r,t_i}^2}}$.

\subsection{Bootstrap.}\ \label{sec:bootstraptheory}
Recall the definition $\boldsymbol{\Upsilon}_i=\frac{1}{\sqrt{2ns'}}\Bigg(\sum_{j \in [i-s'n,i)} \mf Y_j-\sum_{j \in [i,i+s'n)} \mathbf{Y}_j\Bigg)
$ $=(\Upsilon_{1,i},\dots,\Upsilon_{p,i})^\top$ and also recall the definition of $\hat G^{(\ell)}_\text{max}$. In particular,
\begin{align}
\hat{G}^{(\ell)}_{\text{max}}(\mathbf{I})=\max_{\substack{1\leq r\leq p \\ 1\leq j\leq \delta_n \\ t_i\in I_{r}}} \frac{\Bigg|\frac{1}{\sqrt{ns_{r,j}}}\sum_{k =1}^n W\Big(\frac{k/n-t_i}{s_{r,j}}\Big)\Upsilon_{r,k} Z_k^{(\ell)}\Bigg|}{\sqrt{\hat{\sigma}^2_{r,t_i}}}
\end{align}
where $\mf I=(I_1,...,I_p)^\top$ and $I_r$ are sub-intervals of $T_r$, and $(\ell)$ represents the $\ell_{th}$ bootstrap sample. Notice that the $\hat{G}^{(\ell)}_{\text{max}}(\mathbf{I})$'s are conditionally $i.i.d.$ and it suffices to consider the case $\ell=1$. In the following we omit the sup-script $\ell$ for notational brevity. Let $\Delta_n=\max_{r: m_{r,n}\geq 1}\max_i\Delta_{r,i,n}$, where $\Delta_{r,i,n}$ denotes the size of the $i$th jump at dimension $r$ and $m_{r,n}$ denotes the number of jumps at dimension $r$.  
\begin{theorem}\label{Bootstrap}
	Let $\tilde \iota_n=s'+\frac{\log n}{ns'}+\frac{\max_r l_r s'}{\underline s_{\min}}(\sqrt ns'^{3/2})^2+\sqrt{\frac{s'}{\underline s_{\min }}}\log ^5(pn)+\frac{\Delta_n^2m_n ns'^2}{\underline s_{\min}}\wedge \Delta_n^2s'n+\frac{\Delta_nm_nns'^3}{\underline s_{\min}}\wedge \Delta_n(s')^2n+\frac{\Delta_nm_n^{1/2}n^{1/2}s'^{3/2}}{\underline s_{\min}}\log^{5/2}(pn)\wedge \sqrt{\frac{s'}{\underline s_{\min}}}\log^{5/2}(pn)\Delta_n\sqrt{ns'})$, where $m_n=\max_{1\le r\le p}m_{r,n}$ and $l_{r}$ denotes the number of break points at dimension $r$; see Assumption (A1) in Section \ref{sec:apls} for more details.  Assume that 
 $\tilde \iota_n^{1/3}\log^{2/3}p=o(1)$.
	Under conditions of Theorem \ref{Thm1},  we have
	\begin{align}
	&\sup_{\substack{I_r\in T_{r,d},\\1\leq r\leq p,\\x\in \mathbb R}}\big|\p\big(\hat{G}_{\text{max}}(\mathbf{I})\leq x|(\mf Y_i)\big)-\p\big(\sup_{\substack{1\leq r\leq p \\ 1\leq j\leq \delta_n\\ t_i\in I_r}} \frac{|\tilde H^y(t_i,s_{r,j},r)|}{\sqrt{\E\hat{\sigma}_{r,t_i}^2}}\leq x\big)|\notag\\
   &=O_p((n\bar s_{\min})^{c_0-1/4}\sqrt{\log n}+\tilde \iota_n^{1/3}\log^{2/3}p)\notag 
	\end{align}
     for any positive small number $c_0<1/4$. 
\end{theorem}

Together with Theorem \ref{Thm1}, we find that the distribution of the proposed bootstrap well approximates that of $\sup_{\substack{1\leq r\leq p \\ \underline{s}_r\leq s\leq \bar{s}_r\\ t\in I_r}} \frac{|H(t,s,r)|}{\sqrt{\hat{\sigma}_{r,t}^2}}$
at any $I_r\in T_r$, $1\leq r\leq p$, which
validates the procedure of updating $T_{\hat r}^{(a+1)}$ from $T_{\hat r}^{(a)}$ in Algorithm \ref{alg:mainalgo}.
The scale conditions involve $\Delta_n$ due to the difference structure in $\boldsymbol{\Upsilon}_i$.
If $\Delta_n\leq M<\infty$, which is bounded, then the conditions for scales allows
for $s'\asymp n^{-2/3}$, $\underline{s}_{\min}=n^{-1/3+\iota}$ for some $\iota\in(0,1/3)$, $m_n=o(n^\iota)$ and $p=n^{\alpha_0}$ for any $\alpha_0>0$. The scale conditions will be more flexible if $\Delta_n\rightarrow 0$.

\subsection{Estimation Accuracy.}\
Let $\underline \Delta_n=\min_{r: m_{r,n}\geq 1}\min_i\Delta_{r,i,n}$ if there exists a least one jump point.  We omit $n$ in the subscript for the sake of brevity; for example, $\underline \Delta=\underline \Delta_n$.
 
\begin{theorem}\label{thm4}
Under the conditions of Theorem \ref{Bootstrap}, if there exists at least one jump point and sequence $h_n=o(1)$ such that
	\begin{align}
	\bar s_{\max}^2\underline \Delta^{-1}=o(h_n), \underline{\Delta}^{1/2}\bar s_{\max}^{\frac{k+3}{2}}=o(h_n), 
	\frac{\bar s_{\max}\underline \Delta}{n}=o(h_n^2),\label{cond1}\\
	\sqrt{n\underline s_{\min}}\underline \Delta/\log^{5/2}(pn)\rightarrow \infty, \frac{\underline \Delta}{\bar s_{\max}+\frac{1}{n\underline s_{\min}}}\rightarrow \infty,\label{cond2}\\
	\sqrt{\frac{\log^5 (pn)}{n\underline \Delta^2}}=o(h_n), (\frac{\bar s_{\max}^2\log^{7/2}(pn)}{n^{1/2}\bar s_{\min}})^{3/2}=o(h_n)
 \end{align}
 if there is at least one jump point then we have that
	\begin{align}
	\p(\max_{r,v}|\hat d_{r,v}-d_{r,v}|\leq h_n, \hat m_{r,n}=m_{r,n} , \forall r)\rightarrow 1-\alpha
	\end{align}
\end{theorem}
Theorem \ref{thm4} states that all $\hat d_{r,v}$ will deviate from $d_{r,v}$ no more than $h_n$ with probability tending to $1-\alpha$. 
The scale condition allows $\underline \Delta$ to diminish. For example, it allows $p=n^{\alpha_0}$ for any $\alpha_0>0$, $\bar s_{\max}\asymp n^{-1/4}$, $\underline \Delta\asymp n^{-1/4+\iota_0}$ for some $0<\iota_0< 1/4$, $h_n\asymp n^{-1/4-\iota'_0}$ for some $\iota_0'\in (0,\iota_0)$  and $\underline s_{\min}\asymp n^{-1/3+\iota}$ for some $0<\iota \leq 1/12$ defined in the discussion of Theorem \ref{Bootstrap}. Moreover, if $\underline \Delta\geq \eta'>0$ for some positive constant $\eta'$, i.e., the minimal jump size is not varnishing, then if $p=n^{\alpha_0}$ for some $\alpha_0>0$, with the previous choices of $\bar s_{\max}$ and $\underline s_{\min}$, the estimation accuracy of $h_n$ can be $n^{-1/2}(\log^{5/2}n)\iota_n$ for $\iota_n\rightarrow \infty$ arbitrarily slowly. 


\subsection{Second stage high-dimensional local CUSUM.}\label{sec:highdimlocalcusum}
\
\begin{theorem}\label{New.Thm4}
	Let $\iota_n$ be a series diverging arbitrarily slowly.	Assume  $z_n\geq h_n,  \frac{\underline \Delta}{z_n}\rightarrow \infty,$ and conditions of Theorem \ref{thm4} hold, $\frac{nz_n\underline \Delta^2}{\log ^{5}n}\rightarrow \infty$, 
	then we have 
	\begin{align}
	\lim_{n\rightarrow \infty}\p\left(\hat m_{r,n}=m_{r,n}, \max_{r,v}|\tilde d_{r,v}-d_{r,v}|\leq \frac{\iota_n\log^5n}{n\underline  \Delta^2}\right)=  1-\alpha.
	\end{align}
\end{theorem}

\cite{dumbgen1991asymptotic} and \cite{muller1992change} derive the parametric rate $\frac{1}{n\underline \Delta^2}$  for jump detection in low dimensional independent series, which indicates the near optimality of Theorem \ref{New.Thm4} except a factor of logarithm.
We emphasize that this optimality is under the asynchronous jump scenario for high-dimensional and nonstationary time series. 
We compare the estimation error rate of the first stage estimation of Theorem \ref{thm4} and the second stage estimation of Theorem \ref{New.Thm4}, under the scale conditions in the discussion under Theorem \ref{thm4}. If $\underline \Delta\asymp n^{-1/4+\iota_0}$ then the estimation error rate improves from $n^{-1/4-\iota_0'}$ for $\iota_0'\in (0,\iota_0)$ to $n^{-1/2-2\iota_0}(\log^5 n)\iota_n$. If $\underline \Delta\geq \eta$ for some fixed $\eta>0$, then the estimation error rate improves from $n^{-1/2}$ except a factor of logarithm in the first step to $n^{-1}$  except a factor of logarithm. 


\section{Consistency of  AJDN-H.} \label{sec:group}
\begin{assumption}
The probability density function $K(\cdot)$ is Lipschitz continuous, symmetric, and non-zero only on $[-1,1]$.
\label{ass:kernel}
\end{assumption}

\begin{assumption}
    Assume that
\begin{align}
c_1 \leq \min_{1
\leq l \leq m_n}\left\{\max_{1 \leq i \leq K_l^0} |C_{i,l}|/ \min_{1 \leq i \leq K_l^0} |C_{i,l}|\right\} 
\leq \max_{1
\leq l \leq m_n}\left\{\max_{1 \leq i \leq K_l^0} |C_{i,l}|/ \min_{1 \leq i \leq K_l^0}|C_{i,l}| \right\} \leq c_2,
    \label{eq:balance}
\end{align}
where  $c_1$ and $c_2$ are positive constants, 
and 
\begin{align}
  \min_{1 \leq l \leq m_n+1} \min_{1 \leq k_1 < k_2 \leq K_l^0} \int_{t \in \mathcal I_l}| \alpha_{k_1,l}(t) - \alpha_{k_2,l}(t)|^2 dt \geq c_0>0,\label{eq:gap}
\end{align}
where $c_0$ is an absolute constant. 
  \label{ass:gap}
\end{assumption}

Assumption \ref{ass:gap} ensures that the  number in each group is balanced and the integrated squared difference of functions from different groups is uniformly bounded away from $0$.

\begin{assumption}
Summable spatial-temporal covariances
    \begin{align}
        \sum_{j=1}^n \sum_{i=1}^p \max_{1 \leq k \leq n, 1 \leq l \leq p}|\mathrm{Cov} (\epsilon_{i,j}, \epsilon_{l,k})| < \infty.
    \end{align}
    \label{ass:cumulants}
\end{assumption}


\begin{assumption}[Signal strength for hierarchical clustering with jumps]Define 
$\delta_{l}(k_1, k_2) = \int_{t \in \mathcal I_l} |\alpha_{k_1,l}(t)-\alpha_{k_2,l}(t)|\mathrm{d}t$, and $\delta_{n} = \min_{1 \leq l \leq m_n} \min_{1 \leq k_1 \neq k_2 \leq K_l^0}  \delta_{l}(k_1, k_2)$. Assume that
   $\frac{\iota_n\log^5n}{n\underline  \Delta^2} + p^{1/q}(nb )^{-1/2}b^{-1/q} + b  =o(\delta_n)$. \label{ass:HC}
\end{assumption} 
In Assumption \ref{ass:HC}, we allow the integrated absolute difference of functions from different groups to diminish to zero, as long as the magnitude is greater than the localisation rate of change point detection and the estimation error.

\begin{assumption}[Choice of penalty tuning parameter $\chi_n$]
    Assume that 
     \begin{align}
        ((nbp)^{-1/2} + b^3 + \frac{1}{nb})^2 + p^{-1/2} (b^3 + \frac{1}{nb}) + n^{-1} b^{-3/2} p^{-1} (\log n)^2 = o(\chi_n), 
     \end{align}
     and $\chi_n = o(\log b^{-1})$.\label{ass:penalty}
\end{assumption}
Assumption \ref{ass:penalty} gives the upper and lower bound of the penalty parameter $
\chi_n$. Note that in \eqref{eq:sigmal} and \eqref{def:alpha}, we use the group information to pool dimensions with similar mean structures together in estimating and mean functions and $\tilde \sigma_l^2(K)$. Therefore, the estimation accuracy will benefit from the growth of the dimension $p$, and the range of $\chi_n$ will be larger when $p$ increases.

Recall that $m_n$ denotes the total number of distinct jumps in the $p$-dimensional vector $\mf Y_i$, and $K_l^0$ is the number of groups for $\mathcal I_l = (d_l, d_{l+1}]$, where $d_l$ is the $l$-th change point. 
\begin{theorem}\label{thm:information}  
Define the event
\begin{align}
    \mathcal A_n = \{\hat m_{r,n}=m_{r,n}, \max_{r,v}|\tilde d_{r,v}-d_{r,v}|\leq \frac{\iota_n\log^5n}{n\underline  \Delta^2}\}.
\end{align}
Under the conditions of Theorem \ref{New.Thm4}, under Assumptions \ref{ass:kernel}, \ref{ass:gap}, \ref{ass:cumulants},  \ref{ass:HC} and \ref{ass:penalty}, 
we have
\begin{align}
    \mathbb P(\tilde K_l =K_l^0, \{\hat C_{1,l}, \ldots, \hat C_{K_l^0,l}\} = \{C_{1,l}, \ldots, C_{K_l^0,l}\}, \forall 1 \leq l \leq \hat m_n+1 |  \mathcal A_n) \to 1, \ \mathrm{a.s..}
\end{align}
\end{theorem}

\subsection{Improved accuracy of time-varying functions.}\

    In this section, we investigate the integrated MSE
    comparing the proposed method with homogeneity pursuit and the vanilla method that directly smooths the $r$-th dimension. 

    Define the event
\begin{align}
    \mathcal B_n = \{\tilde K_l =K_l^0, \{\hat C_{1,l}, \ldots, \hat C_{K_l^0,l}\} = \{C_{1,l}, \ldots, C_{K_l^0,l}\}, \forall 1 \leq l \leq m_n+1\}.
\end{align}
   On the event that both $\mathcal B_n $ and $\mathcal A_n$ happen, by the calculation in \apxref{sec:IMSE}, we can obtain Integrated MSE of $\beta_r(t)$,
     \begin{align}
        &\int_{0}^1 \E\{|\hat \beta_r (t) -  \beta_r (t)|^2 \mf 1(\mathcal A_n \cap \mathcal B_n)\}\mathrm dt 
        = O(h^4 + \frac{1}{nhp}+  \frac{1}{(nh)^2}). 
        \label{eq:betaimse}
    \end{align}
   
       When $p/ n^{3/2} \to \infty$, 
    the rate is minimized at $h =c_1 n^{-1/3}$ with MISE of order $n^{-4/3}$.  When $p = O(n^{3/2})$, 
    the rate is minimized at $h =c_2 (np)^{-1/5}$ with MISE  of order $(np)^{-4/5}$. Elementary calculation shows that the optimum bandwidths satisfy the conditions in Theorem \ref{thm:information}.

   By contrast, the IMSE of the vanilla component-wise estimation of $\beta_r(t)$ is 
    \begin{align}
        &\int_{0}^1 \E\{|\tilde \beta_r (t) -  \beta_r (t)|^2 \mf 1(\mathcal A_n )\}\mathrm dt = O(h^4 + \frac{1}{nh}), \label{eq:vanilla}
    \end{align}
    whose optimal rate is of order $n^{-4/5}$, achieved at $h =c_3 n^{-1/5}$. Together with Theorem \ref{thm:information}, AJDN-H can improve the estimation accuracy of time-varying functions in the high-dimensional time series.

\section{Simulation studies for detecting, localizing  and testing jumps.}\label{sec:simulationstudy}\
In our simulation studies we study AJDN-H's ability to \LJ{test for}  jumps, achieving correct Type I error rates and good power performance, across a number of simulated data generating processes. \LJ{We also compare AJDN-H's detection and localization performance with a number of modern high-dimensional jump detection methods.} These methods are INSPECT of \cite{wangsamworth2018}, DBLCUSUM of \cite{cho2016}, and LOCLIN of \cite{chen2021inference}. There are numerous other worthy methods to compare AJDN-H to in the literature; however due to page and time constraints we have only compared to these methods as a representative sample. One commonality of these other methods is that they pool information across dimensions, and when detecting a jump at time $t$, they do not explicitly specify in which dimension(s) this jump occurred. This differs from AJDN-H which can explicitly identify in which dimension a jump occurred. Further details about these competing methods, and how they were implemented, can be found in Section \ref{sec:competingimplementation}. For the implementation of AJDN-H the scales and block multiplier bootstrap size were set according to the rule of thumbs in \apxref{apx:selecthyper}, unless otherwise noted. For Type I error simulations 1000 bootstraps were used to calculate critical values, and for each experiment a total of 1000 iterations were run. For power simulations 500 bootstraps and 500 iterations were used.

We note that in our paper when applying these various jump detection methods to real and simulated data, often the assumptions behind them are not met. However, given these methods are representative examples of high-dimensional jump detection methods, it is still useful to examine what can happen to their performance when their assumptions are not met. For our simulation studies we report the average number of jumps detected $\bar{m}$, the percentage of iterations where all jumps were correctly identified with no false positives $\hat{M}_p$, and the average mean absolute deviation (MAD) of the estimated locations compared to the true locations of jumps that have been correctly identified reported on a scale of $[0,n]$ (further details behind MAD calculation are discussed in \apxref{draftfinal:madcalculation}). The proportion of dimensions which experience jumps is denoted $\gamma$. The signal-to-noise ratio, the magnitude of a jump at time $t$ relative to the estimated standard deviation in a small local neighbourhood at time $t$, is denoted by $\mathrm{SN}$.

\subsection{Data generating processes.}\ \label{sec:dgp}
In our simulations we examine the performance of AJDN-H on a variety of data generating processes with and without trends. We generate $\boldsymbol{\epsilon}_i$ from \eqref{eqn:mainmodel} in five different ways
\begin{itemize}
    \item \textbf{(IID)} The errors are independent and identically distributed $\boldsymbol{\epsilon}_i\sim \mathcal{N}_p(0,I)$.
    \item \textbf{(GS)} The errors are globally stationary, following independent AR(1) processes. $\boldsymbol{\epsilon}_i=\mathbf{A}\boldsymbol{\epsilon_{i-1}}+\boldsymbol{\eta}_i$, $\mathbf{A}=\text{diag}(0.25,0.25,\dots,0.25)$ , $\mathbf{A}\in\R^{p\times p}$, $\boldsymbol{\eta}_i\sim N_p(0,I)$.
    \item \textbf{(PS)} The errors are piecewise stationary, following a VMA(3) process where the errors have an equicorrelation structure and a  jump in variance at $t=n/2$.  $\boldsymbol{\epsilon}_i=\boldsymbol{\eta}_i+\mathbf{A}_1\boldsymbol{\eta}_{i-1}+\mathbf{A}_3\boldsymbol{\eta}_{i-3}$, $\mathbf{A}_1=\mathbf{A}_3=\text{diag}(0.5,0.5,\dots,0.5)$, $\mathbf{A}_1,\mathbf{A}_3\in\R^{p\times p}$, that is $\boldsymbol{\eta}_i=\mathbf{B}_i\mathbf{U}_i$, $\mathbf{U}_i\overset{iid}{\sim}\text{Uniform}_p\left[\frac{-\sqrt{12}}{2},\frac{\sqrt{12}}{2}\right]$. For $i\in\{1,2,\dots,n/2\}$ $\mathbf{B}_i\mathbf{B}_i^\top$ is a $p\times p$ matrix with 1 on diagonals and 0.5 on off-diagonals, for $i\in\{n/2+1,\dots,n\}$ $\mathbf{B}_i\mathbf{B}_i^\top$ is equal to 4 on the diagonals and 2 on the off-diagonals.
    \item \textbf{(LS)} the errors are locally stationary, following a VAR(1) model with time varying coefficients where the errors have a Kac-Murdock-Szeg\"{o} correlation structure. Specifically, $\boldsymbol{\epsilon}_i=\frac{\sin\left(2\pi i /n\right)+1}{2}\mathbf{A}_i\boldsymbol{\epsilon}_{i-1}+\boldsymbol{\eta}_i$, $\mathbf{A}_i=\{a_{j,k}\}_{1\leq j,k\leq p}$, $a_{j,k} = 0.25$ for $j=k$ and $j=p-k+1$, and 0 otherwise. $\boldsymbol{\eta}_i=\mathbf{B}_i [\mathbf{Q}_i-3]$, $\mathbf{Q}_i\overset{iid}{\sim}\text{Binomial}_p(10,0.3)$. $\mathbf{B}_i\mathbf{B}_i^\top=\{b_{j,k}\}_{1\leq j,k\leq p}$ with $b_{j,k} = 0.5^{|j-k|}$.
    \item \textbf{(PLS)} the errors are piecewise locally stationary. The data generating process is the same as (LS) with the exception that $\boldsymbol{\eta}_i=c_i\mathbf{B}_i[\mathbf{Q}_i-3]$, where $c_i=1$ for $i \in \{1,\dots,\lceil n/3 \rceil \}\cup \{2\lceil n/3 \rceil,\dots,n\}$ and $c_i=2$ otherwise.
\end{itemize}

Smoothly time varying means were then generated as follows. First, an unscaled mean was generated as $\beta_{r}^{\text{unscaled}}(i/n)=\sin\left(\frac{2\pi i}{n}+\frac{2\pi r}{p} \right) \; i\in \{1,\dots,n\},\; r\in\{1,\dots,p\}.$ Next, the mean was scaled so that the change in the mean at each index $i$ is proportionate to $\hat{\sigma}_{i,r}$, the local estimate of the standard deviation at $t=i/n$ for dimension $r$. In the below results we denote a time series with constant mean as (XYZ) and a time series with a smoothly time varying mean as (XYZT).

\subsection{Implementation details of competing methods.}\label{sec:competingimplementation}\ A high-level overview of the competing methods we compare AJDN-H to, and their implementation details, are given below. Where possible we utilized either the values of hyperparameters suggested by the authors in the paper, or the default values provided in the corresponding R package. We adjust some parameters to ensure the computational cost is manageable. \textbf{INSPECT}: INSPECT uses an optimal projection direction, to project the high-dimensional time series to one dimension, and uses this projection to detect jumps. The implementation used allows for a stationary error process with no dependence between dimensions, and the trend is assumed to be piecewise constant. INSPECT is implemented via the \texttt{Inspectchange point} from \cite{inspectpackage} package in R using 1000 Monte Carlo repetitions for the calculation of the critical value. \textbf{DBLCUSUM}: The test statistic of DBLCUSUM is based on the ordered CUSUM statistics of all dimensions at a given $t$. The error process is assumed to be stationary, and the trend is assumed to be piecewise constant. DBLCUSUM is implemented via the \texttt{hdbinseg} package in R  from \citep{hdbinsegpackage}. We use the version of the test statistic from Section 4.1 of \citep{cho2016}, the same number of bootstrap samples that AJDN-H uses for the calculation of the critical value, and the error process is assumed to have temporal dependence. \textbf{LOCLIN}: LOCLIN uses locally linear estimates of the trend to determine if a jump exists at a given $t$. The error process is assumed to be stationary, and the trend is piecewise smooth. This method was implemented in R based on code provided to us by Weining Wang. Minor modifications were made to the covariance matrix of the Gaussian random vector used in the calculation of the critical value in order to speed up runtime in the high-dimensional setting; in our testing these modifications did not lead to meaningful differences in the Type I Error of the method.  For this method 200 bootstraps were used to calculate the critical value, and the \texttt{KernSmooth} package from \cite{kernsmoothpack} was used to select a bandwidth.

\subsection{Simulations for detecting and localizing jumps.}\label{sec:powersimulatons}
\
Here we compare the ability of AJDN-H to detect jumps relative to other methods' abilities. To do this we consider two scenarios.
\begin{itemize}
	\item \textbf{Scenario 1} $\lceil \gamma p/2 \rceil$ dimensions undergo a jump at $t=0.25$ and a different $\lceil\gamma p/2\rceil$ dimensions undergo a jump at $t=0.75$. The $\gamma$ values tested are $2/p,1/\sqrt{p}$, and 1.
	\item \textbf{Scenario 2} $\lceil\gamma p\rceil$ dimensions undergo jumps at different times, with each of the $\lceil\gamma p\rceil$ dimensions only experiencing a single jump. The first jump occurs at $t=0.2$, the last at $t=0.8$, and all others are evenly spaced in between. The $\gamma$ values tested are $4/p,1/\sqrt{p}$, and $\min\left(\frac{0.6}{2\left(\frac{\log n \log p}{n \Delta_n^2}\right)p},1\right)$ (the minimum is required so that each jump is spaced apart by at least the twice the allowed margin of error). 
\end{itemize}

For all scenarios we use $n=1000$ and test values of 2 and 5 for $\mathrm{SN}$. To make AJDN-H's power results comparable to other methods that do not specify the dimension where  the jump was detected in, we say if a true jump exists in multiple dimensions at $t$, then all jumps detected by AJDN-H within the margin of error  $s_n$ count in total as only one jump having been detected for the purposes of calculating $\bar{m}$. Outside of the margin of error, every unique time where a jump is detected is counted towards the calculation of $\bar{m}$. Due to space constraints we analyze results for a limited set of simulations in this section; further results can be found in \apxref{apx:additionalpower}. 

Table \ref{tab:powersparsity} shows the performance of each algorithm in the (GS) case where jumps are occurring synchronously for different levels of sparsity (i.e. the proportion of dimensions that experience a jump) with stationary noise. Here we see that AJDN-H underperforms LOCLIN and DBLCUSUM in the dense case ($\gamma=1$) as these methods benefit from pooling of information across dimensions; INSPECT on the other hand appears to detect many spurious jumps. In this setting AJDN-H estimates that jumps occur at more times than is actually the case due to estimation error in the jumps that are detected in individual dimensions. We see no improvement in AJDN-H's MAD as $\gamma$ increases, whereas other methods' MAD decreases as the benefit of pooling information across dimensions grows. Interestingly, for AJDN-H the ratio of no false positives  $\hat{M}_p$ is maximized in the $\gamma=1/\sqrt{p}$ setting where a balance is struck between enough dimensions experiencing a jump at a particular time so that AJDN-H detects at least one, and AJDN-H's maximum estimation error being small.

\begin{table}[H]
\caption{\it Simulation results for Scenario 1 with a data generating process of (GS), $\mathrm{SN}=2$ $n=1000,p=100,\alpha=.05$ and varying levels of sparsity $\gamma$.}
\centering
\footnotesize
\scalebox{0.9}{\begin{tabular}{|c|ccc|ccc|ccc|ccc|}
\hline
            & \multicolumn{3}{c|}{AJDN-H}                                                              & \multicolumn{3}{c|}{LOCLIN}                                                            & \multicolumn{3}{c|}{DBLCUSUM}                                                          & \multicolumn{3}{c|}{INSPECT}                                                            \\ \hline
$\gamma$    & \multicolumn{1}{c|}{$\frac{2}{p}$} & \multicolumn{1}{c|}{$\frac{1}{\sqrt{p}}$} & $1$   & \multicolumn{1}{c|}{$\frac{2}{p}$} & \multicolumn{1}{c|}{$\frac{1}{\sqrt{p}}$} & $1$   & \multicolumn{1}{c|}{$\frac{2}{p}$} & \multicolumn{1}{c|}{$\frac{1}{\sqrt{p}}$} & $1$   & \multicolumn{1}{c|}{$\frac{2}{p}$} & \multicolumn{1}{c|}{$\frac{1}{\sqrt{p}}$} & $1$    \\ \hline
$\bar{m}$   & \multicolumn{1}{c|}{1.330}         & \multicolumn{1}{c|}{2.092}                & 2.820 & \multicolumn{1}{c|}{0.772}         & \multicolumn{1}{c|}{1.324}                & 1.902 & \multicolumn{1}{c|}{2.120}         & \multicolumn{1}{c|}{2.112}                & 2.116 & \multicolumn{1}{c|}{161.534}       & \multicolumn{1}{c|}{171.442}              & 175.58 \\ \hline
$\hat{M}_p$ & \multicolumn{1}{c|}{0.392}         & \multicolumn{1}{c|}{0.884}                & 0.414 & \multicolumn{1}{c|}{0.164}         & \multicolumn{1}{c|}{0.420}                & 0.760 & \multicolumn{1}{c|}{0.860}         & \multicolumn{1}{c|}{0.896}                & 0.892 & \multicolumn{1}{c|}{0.000}         & \multicolumn{1}{c|}{0.000}                & 0.000  \\ \hline
MAD         & \multicolumn{1}{c|}{0.887}         & \multicolumn{1}{c|}{0.903}                & 0.917 & \multicolumn{1}{c|}{0.792}         & \multicolumn{1}{c|}{0.455}                & 0.182 & \multicolumn{1}{c|}{0.954}         & \multicolumn{1}{c|}{0.091}                & 0.000 & \multicolumn{1}{c|}{0.729}         & \multicolumn{1}{c|}{0.123}                & 0.000  \\ \hline
\end{tabular}}

\label{tab:powersparsity}
\end{table}

\vspace{-12pt}

\begin{table}[H]
\caption{\it Simulation results for Scenario 2 with $ \mathrm{SN}=5$ $n=1000,p=100,\gamma=\frac{1}{\sqrt{p}}, \alpha=.05$ and varying data generating processes.}
 \fontsize{8.5pt}{11pt}\selectfont
\centering
\scalebox{0.9}{\begin{tabular}{|c|ccc|ccc|ccc|ccc|}
\hline
            & \multicolumn{3}{c|}{AJDN-H}                                          & \multicolumn{3}{c|}{LOCLIN}                                       & \multicolumn{3}{c|}{DBLCUSUM}                                    & \multicolumn{3}{c|}{INSPECT}                                          \\ \hline
DGP         & \multicolumn{1}{c|}{(GS)}   & \multicolumn{1}{c|}{(PS)}   & (PST)  & \multicolumn{1}{c|}{(GS)}  & \multicolumn{1}{c|}{(PS)}   & (PST)  & \multicolumn{1}{c|}{(GS)}  & \multicolumn{1}{c|}{(PS)}  & (PST)  & \multicolumn{1}{c|}{(GS)}    & \multicolumn{1}{c|}{(PS)}    & (PST)   \\ \hline
$\bar{m}$   & \multicolumn{1}{c|}{10.032} & \multicolumn{1}{c|}{10.036} & 10.044 & \multicolumn{1}{c|}{8.066} & \multicolumn{1}{c|}{15.562} & 16.282 & \multicolumn{1}{c|}{8.992} & \multicolumn{1}{c|}{9.722} & 10.608 & \multicolumn{1}{c|}{174.106} & \multicolumn{1}{c|}{252.602} & 253.750 \\ \hline
$\hat{M}_p$ & \multicolumn{1}{c|}{0.962}  & \multicolumn{1}{c|}{0.930}  & 0.932  & \multicolumn{1}{c|}{0.460} & \multicolumn{1}{c|}{0.000}  & 0.000  & \multicolumn{1}{c|}{0.252} & \multicolumn{1}{c|}{0.040} & 0.006  & \multicolumn{1}{c|}{0.000}   & \multicolumn{1}{c|}{0.000}   & 0.000   \\ \hline
MAD         & \multicolumn{1}{c|}{0.022}  & \multicolumn{1}{c|}{0.031}  & 0.040  & \multicolumn{1}{c|}{0.056} & \multicolumn{1}{c|}{0.125}  & 0.128  & \multicolumn{1}{c|}{0.024} & \multicolumn{1}{c|}{0.031} & 0.042  & \multicolumn{1}{c|}{0.115}   & \multicolumn{1}{c|}{0.107}   & 0.172   \\ \hline
\end{tabular}}
\label{tab:powernonstationary}
\end{table}

\vspace{-12pt}

Next we examine results from Scenario 2 where there is no benefit from pooling information across dimensions, and vary the data generating process to introduce progressively more nonstationarity. Table \ref{tab:powernonstationary} summarizes results in the relatively sparse jump case ($\gamma=1/\sqrt{p}$) where the total number of jumps is 10. In the high signal-to-noise ratio case of $ \mathrm{SN}=5$, AJDN-H performs best in the (GS) setting where $\hat{M}_p\approx 1-\alpha$ and slightly worse in (PS) and (PST) where MAD increases and $\hat{M}_p$ correspondingly decreases. The performances of other methods in the (GS) case are inferior to AJDN-H because of the asynchronicity of the jumps.  We also see a deterioration of performance (particularly in $\hat{M}_p$) of the latter three methods as the error process changes from (GS) to (PS) and a increase in spurious jumps detected as a trend is added in the (PST) setting.

Finally, we present results in Table \ref{tab:powerasynchronous} demonstrating the performance of each algorithm in Scenario 2 with $\gamma=1$ where a large number of jumps are occurring asynchronously. This setting naturally lends itself well to the methodology of AJDN-H.
\begin{table}[H]
\caption{\it Simulation results for Scenario 2 with $\mathrm{SN} = 5$, $n=1000,p=100,\gamma=1, \alpha=.05$ and varying data generating processes.}
\fontsize{8pt}{11pt}\selectfont
\centering
\scalebox{0.9}{\begin{tabular}{|c|ccc|ccc|ccc|ccc|}
\hline
            & \multicolumn{3}{c|}{AJDN}                                           & \multicolumn{3}{c|}{LOCLIN}                                       & \multicolumn{3}{c|}{DBLCUSUM}                                      & \multicolumn{3}{c|}{INSPECT}                                         \\ \hline
DGP         & \multicolumn{1}{c|}{(IID)}   & \multicolumn{1}{c|}{(PLS)}  & (PLST) & \multicolumn{1}{c|}{(IID)} & \multicolumn{1}{c|}{(PLS)}  & (PLST) & \multicolumn{1}{c|}{(IID)}  & \multicolumn{1}{c|}{(PLS)}  & (PLST) & \multicolumn{1}{c|}{(IID)}  & \multicolumn{1}{c|}{(PLS)}   & (PLST)  \\ \hline
$\bar{m}$   & \multicolumn{1}{c|}{100.008} & \multicolumn{1}{c|}{99.286} & 99.068 & \multicolumn{1}{c|}{6.790} & \multicolumn{1}{c|}{10.436} & 12.994 & \multicolumn{1}{c|}{13.322} & \multicolumn{1}{c|}{13.584} & 13.890 & \multicolumn{1}{c|}{76.004} & \multicolumn{1}{c|}{136.592} & 143.764 \\ \hline
$\hat{M}_p$ & \multicolumn{1}{c|}{0.986}   & \multicolumn{1}{c|}{0.766}  & 0.758  & \multicolumn{1}{c|}{0.000} & \multicolumn{1}{c|}{0.000}  & 0.000  & \multicolumn{1}{c|}{0.000}  & \multicolumn{1}{c|}{0.000}  & 0.000  & \multicolumn{1}{c|}{0.000}  & \multicolumn{1}{c|}{0.000}   & 0.000   \\ \hline
MAD         & \multicolumn{1}{c|}{0.020}   & \multicolumn{1}{c|}{0.032}  & 0.036  & \multicolumn{1}{c|}{0.272} & \multicolumn{1}{c|}{0.251}  & 0.189  & \multicolumn{1}{c|}{0.288}  & \multicolumn{1}{c|}{0.242}  & 0.207  & \multicolumn{1}{c|}{0.255}  & \multicolumn{1}{c|}{0.115}   & 0.119   \\ \hline
\end{tabular}}
\label{tab:powerasynchronous}
\end{table}

Here 100 jumps are present. First we examine the IID case, and hence fall within the assumptions on the error process for all methods. Here we see AJDN-H strongly outperform all other methods, perfectly identifying all jumps 98.6\% of the time.  
INSPECT moderately underestimated the number of jumps, and for the jumps it does detect, has a significantly higher MAD than AJDN-H. In the (PLS) and (PLST) setting AJDN-H performs strongly, perfectly identifying all jumps approximately 75\% of the time.

\subsection{Simulations for testing jumps.}\ 
AJDN-H  can be used as a multiscale test to determine if any jumps exist within a high-dimensional time series. The null hypothesis is that each time series has a smooth trend with no jumps. Consider the following model
\begin{align}
	\mathbf{Y}_{i}&=\boldsymbol{\beta}(i/n)+\Delta \sigma_{1,i/n=.5}\mathbf{1}_{1}(0.5<i/n\leq 1) + \boldsymbol{\epsilon}_{i}	 \label{type1model}
\end{align}
where $\boldsymbol{\beta}(i/n)$ is either constant in the no trend case, or a vector of mean functions generated according to Section \ref{sec:dgp} in the trend case. $\mathbf{1}_{1}(0.5<t\leq 1)$ represents a jump in dimension one occurring immediately after $i/n=.5$, $\sigma_{1,i/n=.5}$ is the standard deviation of the first dimension's error process at $i/n=.5$, and $\boldsymbol{\epsilon_{i}}$ is an error process. Using the rule of thumb for $(\underline{s},\bar{s})$, and the process for selecting $s'$, laid out in \apxref{apx:selecthyper}, we observe the Type I errors (i.e. rejection rates when $\Delta=0$) produced by our test in Table \ref{draftsimsfinal:tab:type1errors}. Table \ref{draftfinal:tab:type1compare} shows the false positive rates of the various jump detection methods we are comparing AJDN-H to across a variety of data generating processes. We see in this table that AJDN-H is still able to control the Type I error successfully even for very complex data generating processes such as (PLST), whereas the other methods struggle.  \apxref{apx:mjpdhdpowercurve} of the appendix also contains the simulated power performance for AJDN-H with different values of $\Delta$.


\begin{table}[H]
\caption{\it Type I errors produced by different jump detection methods $n=2000$, $p=100$ $\alpha=0.05$.}
\footnotesize
\centering
\scalebox{1}{
\begin{tabular}{ccccccccccc}
  \hline
 & (IID) & (GS) & (PS) & (LS) & (PLS) & (IIDT) & (GST) & (PST) & (LST) & (PLST) \\   \hline
  AJDN-H & 0.033 & 0.077 & 0.051 & 0.009 & 0.015 & 0.036 & 0.058 & 0.047 & 0.013 & 0.020 \\ 
  LOCLIN & 0.061 & 0.110 & 1.000 & 0.951 & 1.000 & 0.058 & 0.118 & 1.000 & 0.949 & 1.000 \\ 
  DBLCUSUM & 0.051 & 0.037 & 0.372 & 0.154 & 0.000 & 1.000 & 1.000 & 1.000 & 1.000 & 1.000 \\ 
  INSPECT & 0.001 & 0.999 & 1.000 & 0.946 & 0.984 & 1.000 & 1.000 & 1.000 & 1.000 & 1.000 \\ 
   \hline
\end{tabular}
} 
\label{draftfinal:tab:type1compare}
\end{table}

\section{Simulation studies for estimating mean functions}\label{sec:simgrou}
\subsection{Selection of tuning parameters.}\label{sec:b}\ 
For combining the jumps that are close to each other, we set $s_n = (
\log n)^5/n $, which is of the same magnitude of the localization rate in Theorem \ref{New.Thm4}.
By similar arguments in 
the proof of Theorem \ref{thm:information}, we can obtain Integrated MSE of $\alpha_{k,l}(t)$, 
    \begin{align}
        \int_{0}^1 \E|\{\check \alpha_{k,l} (t) -  \alpha_{k,l} (t)\}\mf 1(\mathcal A_n \cap \mathcal B_n)|^2 \mathrm dt
        & = O(b^4 + \frac{1}{nbp}+  \frac{1}{(nb)^2}).\notag
    \end{align}
       When $p/ n^{3/2} \to \infty$, 
    the rate is minimized at $b =c n^{-1/3}$ with MISE of order $n^{-4/3}$.  When $p = O(n^{3/2})$, 
    the rate is minimized at $b =c (np)^{-1/5}$ with MISE  of order $(np)^{-4/5}$. In the simulation, we set $c = 0.35$.
In the simulation, we use $h = 0.65 n^{-1/5}$ for the vanilla method and the proposed method.
We set the penalty parameter $\chi_n = \sqrt{\log (nb) b/n}$, which satisfy the condition in   Assumption \ref{ass:penalty}. 


\subsection{Simulation Settings.}\
In our simulations we examine the estimation accuracy of trends for AJDN-H on a variety of data generating processes with and without specific homogeneity structures. We generate $\boldsymbol{\epsilon}_i$ from \eqref{eqn:mainmodel} in \textbf{(GS)} and \textbf{(LS)}, see Section \ref{sec:dgp}.



We consider mean functions of $\boldsymbol \epsilon_i$ in the following two cases: 
\begin{enumerate}
    
\item \textbf{Data without specific homogeneity structures. } 
Smoothly time varying means were generated as follows. First, an unscaled mean was generated as \begin{align}
    \gamma_{r}(i/n)=\sin\left(\frac{\pi i}{n} \right) +2\pi r\;, i\in \{1,\dots,n\},\; r\in\{1,\dots,p\}. \notag 
\end{align}
 Next, the mean was scaled   proportionate to the local estimate of the standard deviation as in Section \ref{sec:dgp}.

\item \textbf{Data with  homogeneity structures. }  For the case with homogeneity structures, we \LJ{directly} consider the following alternating functions for $j=1,\ldots, \lfloor p/3 \rfloor$, 
\begin{align}
    \gamma_{3j+1} (i/n) = (i/n-1/2)^2, \quad  \gamma_{3j+2} (i/n) =  (i/n-1/2), \quad \gamma_{3j} (i/n) = \sin(\pi i/n), \notag
\end{align}
which exhibit homogeneity structures.
\end{enumerate}

Based on the aforementioned mean functions, we consider the asychronous jumps.
We set the jump size $\Delta=5$ and let
$\lceil p/2 \rceil$ dimensions randomly undergo a jump at $t=0.25$ and a different $\lceil p/2\rceil$ dimensions randomly undergo a jump at $t=0.75$, for $1 \leq r \leq p$,
\begin{align}
\beta_r (i/n) = \gamma_r(i/n) + \Delta \mf 1(i/n > 0.25, \textrm{jump at} \  0.25 ) + \Delta \mf 1(i/n > 0.75, \textrm{jump at} \  0.75 ).\notag 
\end{align}

\begin{table}[ht]
 \fontsize{8.5pt}{10pt}\selectfont
\begin{minipage}{0.45\textwidth}
\begin{tabular}{cccccc}
 \hline
 \multicolumn{2}{c}{$n=$} &\multicolumn{2}{c}{500}&\multicolumn{2}{c}{1000}\\
 \hline
dgp & K & IMSE1 & IMSE2 & IMSE1 & IMSE2 \\ 
\hline
\multirow{3}{*}{GS}&0 & 4.70 & 4.70 & 3.00 & 3.00 \\ 
 & 2 & 3.57 & 3.57 & 1.96 & 1.96 \\ 
 & 10 & 4.09 & 4.09 & 2.18 & 2.18 \\ \hline
\multirow{3}{*}{LS} & 0 & 7.16 & 7.16 & 4.62 & 4.62 \\ 
 & 2 & 6.81 & 6.81 & 3.52 & 3.52 \\ 
 & 10 & 9.63 & 9.63 & 3.76 & 3.76 \\ 
  \hline
\end{tabular}\caption{IMSE averaged by dimension $p=200$ of the vanilla method (IMSE1) and the proposed method (IMSE2) without  homogeneity structures.}
\end{minipage}
\hfill
\begin{minipage}{0.45\textwidth}
\begin{tabular}{cccccc}
 \hline
 \multicolumn{2}{c}{$n=$} &\multicolumn{2}{c}{500}&\multicolumn{2}{c}{1000}\\
 \hline
dgp & K & IMSE1 & IMSE2 & IMSE1 & IMSE2 \\ 
\hline
\multirow{3}{*}{GS} & 0 & 4.65 & 2.04 & 2.94 & 1.72 \\ 
 & 2 & 3.47 & 0.87 & 1.89 & 0.64 \\ 
 & 10 & 4.00 & 1.41 & 2.10 & 0.85 \\ 
 \hline
\multirow{3}{*}{LS} &0 & 7.00 & 1.92 & 4.41 & 1.80 \\ 
 & 2 & 6.62 & 3.18 & 3.36 & 0.72 \\ 
 & 10 & 9.41 & 7.87 & 3.62 & 0.98 \\ 
 \hline
\end{tabular}
\caption{IMSE averaged by dimension $p=200$ of the vanilla method (IMSE1) and the proposed method (IMSE2) with  homogeneity structures.}\end{minipage}
\end{table}

\section{Real data analysis. \ }\label{sec:stocksanalysis} \ 
Closing daily stock prices adjusted for dividends for the Nasdaq-100 stocks were collected from 2021-01-01 to 2022-12-31 using the \texttt{quantmod} package in R from \cite{quantmodpackage}.
 In order to only detect changes in the values of these companies, as opposed to changes in the overall market, we divide each stock price by the value of the price of QQQ, an exchange-traded fund which tracks the Nasdaq index. By transforming the data in this fashion, we are left with time series which appear to follow a smooth trend plus noise model in between jumps. Including data for all trading days over this time period for these stocks produces an $n=503,\; p=96$ dataset. 

First, for better illustration, visualization and analyzing the event behind the jumps, we  detect times at which the values of FAANG stocks (Facebook, Amazon, Apple, Netflix, and Google)  undergo a significant change.

To select both $s'$ and $\{(\underline{s},\bar{s})\}_{r=1}^p$ we select the combination of these parameters that minimize the average penalized-BIC across all dimensions. For computational reasons we set $(\underline{s}_r,\bar{s}_r)=(\underline{s},\bar{s})$ for all $r=1,\dots,p$. Given that these companies report financial results every quarter throughout the year, we test a range of candidate $\bar{s}$ of approximately half of a quarter ($\bar{s}\approx .0625$), and test all $ns'$ from 1 to $ns'_\text{max}$. Using this criterion, the hyperparameters selected were $s'=0.0020$ and $(\underline{s},\bar{s})=(.027,.044)$.

\begin{figure}
	\centering
	\includegraphics[width=13cm]{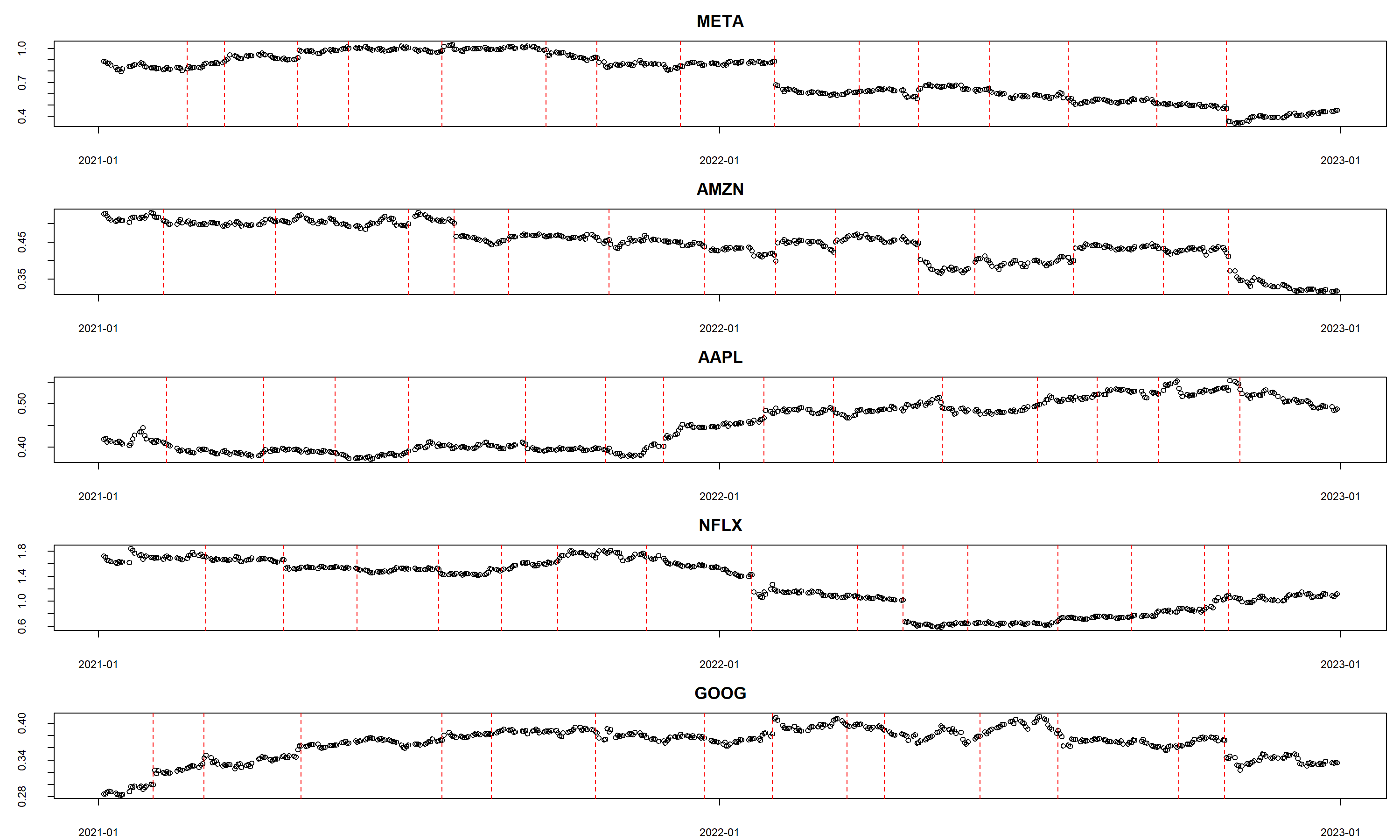}
	\caption{\it Jumps identified by AJDN-H across 5 technology stocks in 2021 and 2022.}
	\label{draftfinal:fig:stockscpdetected}
\end{figure}

We see from Figure \ref{draftfinal:fig:stockscpdetected} that there is a mixture of times where many stocks experience a jump around the same time, and some where only one stock experiences a jump. The shared times tend to coincide with when these companies would report their quarterly earnings. A total of 72 jumps are detected across all 5 stocks. To examine the validity of the jumps detected by AJDN-H, we examine a representative dimension, AMZN, and attempt to explain if there were underlying events which can explain the detected jumps.

AJDN-H detects a total of 14 jumps for AMZN stock over the course of 2021 and 2022. A visual inspection of Figure \ref{draftfinal:fig:stockscpdetected} would suggest that most of the major jumps were successfully identified. Utilizing ProQuest and Google News we are able to identify events which were the likely driver of these jumps on 13 of the 14 days, 7 of which are quarterly earnings related announcements. News articles describing the events likely driving these jumps can be found in Table 7, and their sources can be found in Table 8 in the online supplementary material.

Second, to study the effect of homogeneity pursuit step in AJDN-H when the dimension is large, we analyze all the Nasdaq-100 stock prices that are available with $p=96$. We first combining the jumps that are close to each other, i.e., their distance is smaller than $s_n=(\log n)^5/n$. We obtain 10 jumps. 
Then, we employed the proposed homogeneity pursuit step of AJDN-H. 
We report the 35-60 of clusters over the period. The relative large number of cluster ensures the estimtation accuracy via avoiding under splitting. 

\begin{figure}
    \centering
    \includegraphics[width=0.5\linewidth]{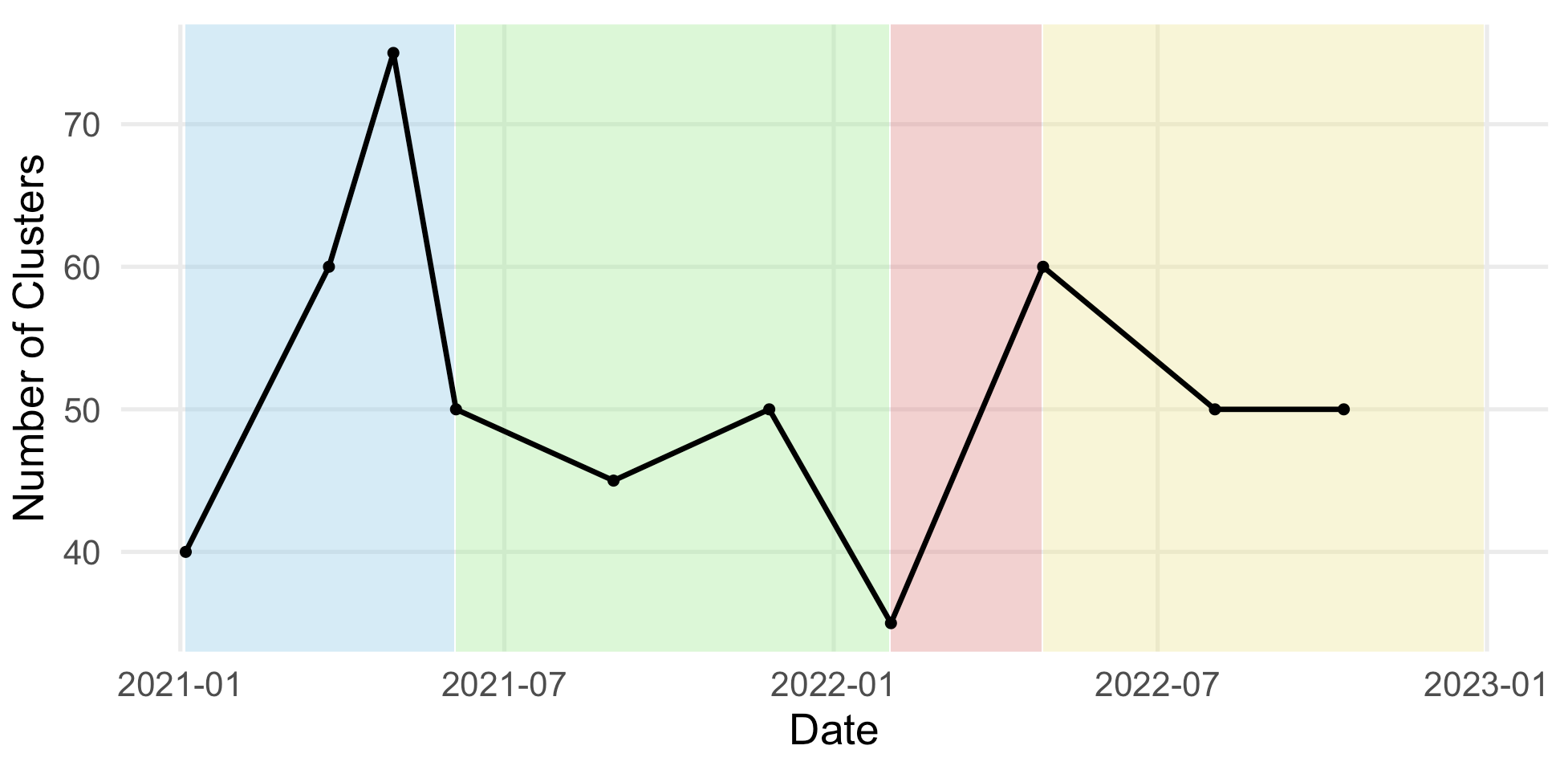}
    \caption{Number of clusters   from 2021-01-01 to 2022-12-31 of  Nasdaq-100.}
    \label{fig:clusters}
\end{figure}

As shown in Figure \ref{fig:clusters}, the variation in cluster counts aligns well with major macroeconomic and geopolitical events shaping market behavior during this period. The sharp rise in clusters from early to mid-2021 coincides with the global reopening phase after COVID-19 lockdowns, vaccine rollouts, and strong fiscal stimulus, which led to sector rotations and increased dispersion across assets. The subsequent decline into late 2021 reflects a more synchronized bull market as liquidity remained abundant. The pronounced drop to the lowest cluster level in early 2022 matches the onset of the Russian invasion of Ukraine and escalating inflation concerns, which drove a unified risk-off sentiment across markets. The rebound in clusters through mid-2022 corresponds to aggressive monetary tightening by central banks such as the Federal Reserve, creating divergence between sectors and asset classes. Finally, the stabilization in late 2022 suggests markets adjusting to a “new normal” of higher interest rates and persistent uncertainty, with fewer dominant shocks but no clear directional consensus. The comparison of estimated mean functions can be found in \apxref{apx:mean}.

 \begin{funding}  \
 Weichi Wu is the corresponding author and is supported by NSFC 12271287. This work was partially supported by DFG Research unit 5381 Mathematical
Statistics in the Information Age (Project number 460867398) and by TRR 391 Spatio-temporal
Statistics for the Transition of Energy and Transport (Project number 520388526) funded by the
Deutsche Forschungsgemeinschaft (DFG, German Research Foundation).
\end{funding}
\section*{Supplementary material.\  }
The supplement contains proofs, addition results and details for simulation  and more results for the data analysis.


\bibliographystyle{imsart-nameyear}
\bibliography{main}

\begin{thebibliography}{40}

\bibitem[\protect\citeauthoryear{Abdollahi, Junttila and
  Lehkonen}{2024}]{ABDOLLAHI2024102004}
\begin{barticle}[author]
\bauthor{\bsnm{Abdollahi},~\bfnm{Hooman}\binits{H.}},
  \bauthor{\bsnm{Junttila},~\bfnm{Juha-Pekka}\binits{J.-P.}} \AND
  \bauthor{\bsnm{Lehkonen},~\bfnm{Heikki}\binits{H.}}
(\byear{2024}).
\btitle{Clustering asset markets based on volatility connectedness to political
  news}.
\bjournal{Journal of International Financial Markets, Institutions and Money}
\bvolume{93}
\bpages{102004}.
\bdoi{https://doi.org/10.1016/j.intfin.2024.102004}
\end{barticle}
\endbibitem

\bibitem[\protect\citeauthoryear{Bai and Wu}{2025}]{bai2025}
\begin{barticle}[author]
\bauthor{\bsnm{Bai},~\bfnm{Lin}\binits{L.}} \AND
  \bauthor{\bsnm{Wu},~\bfnm{Wei}\binits{W.}}
(\byear{2025}).
\btitle{Uniform variance reduced simultaneous inference of time-varying
  correlation networks}.
\bjournal{IEEE Transactions on Information Theory}
\bpages{1--1}.
\bdoi{10.1109/TIT.2025.3613143}
\end{barticle}
\endbibitem

\bibitem[\protect\citeauthoryear{Chen, Wang and Wu}{2022}]{chen2021inference}
\begin{barticle}[author]
\bauthor{\bsnm{Chen},~\bfnm{Lijun}\binits{L.}},
  \bauthor{\bsnm{Wang},~\bfnm{Weiguo}\binits{W.}} \AND
  \bauthor{\bsnm{Wu},~\bfnm{Wei~Biao}\binits{W.~B.}}
(\byear{2022}).
\btitle{Inference of breakpoints in high-dimensional time series}.
\bjournal{Journal of the American Statistical Association}
\bvolume{117}
\bpages{1951--1963}.
\end{barticle}
\endbibitem

\bibitem[\protect\citeauthoryear{Chernozhukov, Chetverikov and
  Kato}{2017}]{chernozhukov2017central}
\begin{barticle}[author]
\bauthor{\bsnm{Chernozhukov},~\bfnm{Victor}\binits{V.}},
  \bauthor{\bsnm{Chetverikov},~\bfnm{Denis}\binits{D.}} \AND
  \bauthor{\bsnm{Kato},~\bfnm{Kengo}\binits{K.}}
(\byear{2017}).
\btitle{Central limit theorems and bootstrap in high dimensions}.
\bjournal{The Annals of Probability}
\bvolume{45}
\bpages{2309--2352}.
\end{barticle}
\endbibitem

\bibitem[\protect\citeauthoryear{Cho}{2016}]{cho2016}
\begin{barticle}[author]
\bauthor{\bsnm{Cho},~\bfnm{Haeran}\binits{H.}}
(\byear{2016}).
\btitle{Change-point detection in panel data via double CUSUM statistic}.
\bjournal{Electronic Journal of Statistics}
\bvolume{10}
\bpages{2000--2038}.
\end{barticle}
\endbibitem

\bibitem[\protect\citeauthoryear{Cho and Fryzlewicz}{2014}]{cho2015}
\begin{barticle}[author]
\bauthor{\bsnm{Cho},~\bfnm{Haeran}\binits{H.}} \AND
  \bauthor{\bsnm{Fryzlewicz},~\bfnm{Piotr}\binits{P.}}
(\byear{2014}).
\btitle{Multiple-Change-Point Detection for High Dimensional Time Series via
  Sparsified Binary Segmentation}.
\bjournal{Journal of the Royal Statistical Society Series B: Statistical
  Methodology}
\bvolume{77}
\bpages{475-507}.
\bdoi{10.1111/rssb.12079}
\end{barticle}
\endbibitem

\bibitem[\protect\citeauthoryear{Cho and Fryzlewicz}{2018}]{hdbinsegpackage}
\begin{bmisc}[author]
\bauthor{\bsnm{Cho},~\bfnm{Haeran}\binits{H.}} \AND
  \bauthor{\bsnm{Fryzlewicz},~\bfnm{Piotr}\binits{P.}}
(\byear{2018}).
\btitle{{hdbinseg: Change-Point Analysis of High-Dimensional Time Series via
  Binary Segmentation}}.
\bnote{R package version 1.0.1}.
\end{bmisc}
\endbibitem

\bibitem[\protect\citeauthoryear{Dette and G{\"o}smann}{2018}]{2018dette}
\begin{barticle}[author]
\bauthor{\bsnm{Dette},~\bfnm{Holger}\binits{H.}} \AND
  \bauthor{\bsnm{G{\"o}smann},~\bfnm{Jonas}\binits{J.}}
(\byear{2018}).
\btitle{Relevant change points in high dimensional time series}.
\bjournal{Electronic Journal of Statistics}
\bvolume{12}
\bpages{2578--2636}.
\end{barticle}
\endbibitem

\bibitem[\protect\citeauthoryear{Dette and Wu}{2024}]{dette2021confidence}
\begin{barticle}[author]
\bauthor{\bsnm{Dette},~\bfnm{Holger}\binits{H.}} \AND
  \bauthor{\bsnm{Wu},~\bfnm{Wei}\binits{W.}}
(\byear{2024}).
\btitle{Confidence surfaces for the mean of locally stationary functional time
  series}.
\bjournal{Statistica Sinica}.
\bnote{To appear, doi:10.5705/ss.202023.0150}.
\end{barticle}
\endbibitem

\bibitem[\protect\citeauthoryear{D{\"u}mbgen}{1991}]{dumbgen1991asymptotic}
\begin{barticle}[author]
\bauthor{\bsnm{D{\"u}mbgen},~\bfnm{Lutz}\binits{L.}}
(\byear{1991}).
\btitle{The asymptotic behavior of some nonparametric change-point estimators}.
\bjournal{The Annals of Statistics}
\bvolume{19}
\bpages{1471--1495}.
\end{barticle}
\endbibitem

\bibitem[\protect\citeauthoryear{Enikeeva and Harchaoui}{2019}]{enikeeva2019}
\begin{barticle}[author]
\bauthor{\bsnm{Enikeeva},~\bfnm{Fanny}\binits{F.}} \AND
  \bauthor{\bsnm{Harchaoui},~\bfnm{Zaid}\binits{Z.}}
(\byear{2019}).
\btitle{High-dimensional change-point detection under sparse alternatives}.
\bjournal{The Annals of Statistics}
\bvolume{47}
\bpages{2051--2079}.
\end{barticle}
\endbibitem

\bibitem[\protect\citeauthoryear{Gao, Gijbels and
  Van~Bellegem}{2008}]{gao2008nonparametric}
\begin{barticle}[author]
\bauthor{\bsnm{Gao},~\bfnm{Jiti}\binits{J.}},
  \bauthor{\bsnm{Gijbels},~\bfnm{Irene}\binits{I.}} \AND
  \bauthor{\bsnm{Van~Bellegem},~\bfnm{St{\'e}phane}\binits{S.}}
(\byear{2008}).
\btitle{Nonparametric simultaneous testing for structural breaks}.
\bjournal{Journal of Econometrics}
\bvolume{143}
\bpages{123--142}.
\end{barticle}
\endbibitem

\bibitem[\protect\citeauthoryear{Grundy, Killick and
  Mihaylov}{2020}]{grundy2020high}
\begin{barticle}[author]
\bauthor{\bsnm{Grundy},~\bfnm{Tom}\binits{T.}},
  \bauthor{\bsnm{Killick},~\bfnm{Rebecca}\binits{R.}} \AND
  \bauthor{\bsnm{Mihaylov},~\bfnm{George}\binits{G.}}
(\byear{2020}).
\btitle{High-dimensional changepoint detection via a geometrically inspired
  mapping}.
\bjournal{Statistics and Computing}
\bvolume{30}
\bpages{1155--1166}.
\end{barticle}
\endbibitem

\bibitem[\protect\citeauthoryear{Han, Wu and Zhang}{2025}]{han2025new}
\begin{barticle}[author]
\bauthor{\bsnm{Han},~\bfnm{Yang}\binits{Y.}},
  \bauthor{\bsnm{Wu},~\bfnm{Weichi}\binits{W.}} \AND
  \bauthor{\bsnm{Zhang},~\bfnm{Wenyang}\binits{W.}}
(\byear{2025}).
\btitle{A New Approach for Homogeneity Pursuit in Short Panel Data Analysis}.
\bjournal{Journal of the American Statistical Association}
\bpages{1--11}.
\end{barticle}
\endbibitem

\bibitem[\protect\citeauthoryear{Hardy}{2001}]{hardy2001regime}
\begin{barticle}[author]
\bauthor{\bsnm{Hardy},~\bfnm{Mary~R.}\binits{M.~R.}}
(\byear{2001}).
\btitle{A regime-switching model of long-term stock returns}.
\bjournal{North American Actuarial Journal}
\bvolume{5}
\bpages{41--53}.
\end{barticle}
\endbibitem

\bibitem[\protect\citeauthoryear{Jirak}{2015}]{jirak2015uniform}
\begin{barticle}[author]
\bauthor{\bsnm{Jirak},~\bfnm{Milan}\binits{M.}}
(\byear{2015}).
\btitle{Uniform change point tests in high dimension}.
\bjournal{The Annals of Statistics}
\bvolume{43}
\bpages{2451--2483}.
\end{barticle}
\endbibitem

\bibitem[\protect\citeauthoryear{Ke, Fan and Wu}{2015}]{ke2015homogeneity}
\begin{barticle}[author]
\bauthor{\bsnm{Ke},~\bfnm{Zheng~Tracy}\binits{Z.~T.}},
  \bauthor{\bsnm{Fan},~\bfnm{Jianqing}\binits{J.}} \AND
  \bauthor{\bsnm{Wu},~\bfnm{Yichao}\binits{Y.}}
(\byear{2015}).
\btitle{Homogeneity pursuit}.
\bjournal{Journal of the American Statistical Association}
\bvolume{110}
\bpages{175--194}.
\end{barticle}
\endbibitem

\bibitem[\protect\citeauthoryear{Ke, Li and Zhang}{2016}]{ke2016}
\begin{barticle}[author]
\bauthor{\bsnm{Ke},~\bfnm{Yuan}\binits{Y.}},
  \bauthor{\bsnm{Li},~\bfnm{Jialiang}\binits{J.}} \AND
  \bauthor{\bsnm{Zhang},~\bfnm{Wenyang}\binits{W.}}
(\byear{2016}).
\btitle{{Structure identification in panel data analysis}}.
\bjournal{The Annals of Statistics}
\bvolume{44}
\bpages{1193 -- 1233}.
\bdoi{10.1214/15-AOS1403}
\end{barticle}
\endbibitem

\bibitem[\protect\citeauthoryear{Kong et~al.}{2017}]{Kong2017}
\begin{barticle}[author]
\bauthor{\bsnm{Kong},~\bfnm{Yinfei}\binits{Y.}},
  \bauthor{\bsnm{Li},~\bfnm{Daoji}\binits{D.}},
  \bauthor{\bsnm{Fan},~\bfnm{Yingying}\binits{Y.}} \AND
  \bauthor{\bsnm{Lv},~\bfnm{Jinchi}\binits{J.}}
(\byear{2017}).
\btitle{{Interaction pursuit in high-dimensional multi-response regression via
  distance correlation}}.
\bjournal{The Annals of Statistics}
\bvolume{45}
\bpages{897 -- 922}.
\bdoi{10.1214/16-AOS1474}
\end{barticle}
\endbibitem

\bibitem[\protect\citeauthoryear{Li and Zhao}{2019}]{li2019time}
\begin{barticle}[author]
\bauthor{\bsnm{Li},~\bfnm{Xuefeng}\binits{X.}} \AND
  \bauthor{\bsnm{Zhao},~\bfnm{Zhijie}\binits{Z.}}
(\byear{2019}).
\btitle{A time varying approach to the stock return--inflation puzzle}.
\bjournal{Journal of the Royal Statistical Society: Series C (Applied
  Statistics)}
\bvolume{68}
\bpages{1509--1528}.
\end{barticle}
\endbibitem

\bibitem[\protect\citeauthoryear{Li et~al.}{2022}]{li2022ell}
\begin{barticle}[author]
\bauthor{\bsnm{Li},~\bfnm{Jing}\binits{J.}},
  \bauthor{\bsnm{Chen},~\bfnm{Lijun}\binits{L.}},
  \bauthor{\bsnm{Wang},~\bfnm{Weiguo}\binits{W.}} \AND
  \bauthor{\bsnm{Wu},~\bfnm{Wei~Biao}\binits{W.~B.}}
(\byear{2022}).
\btitle{$\ell^{2}$ inference for change points in high-dimensional time series
  via a two-way MOSUM}.
\bjournal{arXiv preprint arXiv:2208.13074}.
\end{barticle}
\endbibitem

\bibitem[\protect\citeauthoryear{Liu, Zhang and Liu}{2022}]{liu2022high}
\begin{barticle}[author]
\bauthor{\bsnm{Liu},~\bfnm{Bing}\binits{B.}},
  \bauthor{\bsnm{Zhang},~\bfnm{Xiangyu}\binits{X.}} \AND
  \bauthor{\bsnm{Liu},~\bfnm{Yao}\binits{Y.}}
(\byear{2022}).
\btitle{High dimensional change point inference: Recent developments and
  extensions}.
\bjournal{Journal of Multivariate Analysis}
\bvolume{188}
\bpages{104833}.
\end{barticle}
\endbibitem

\bibitem[\protect\citeauthoryear{M{\"u}ller}{1992}]{muller1992change}
\begin{barticle}[author]
\bauthor{\bsnm{M{\"u}ller},~\bfnm{Hans-Georg}\binits{H.-G.}}
(\byear{1992}).
\btitle{Change-points in nonparametric regression analysis}.
\bjournal{The Annals of Statistics}
\bpages{737--761}.
\end{barticle}
\endbibitem

\bibitem[\protect\citeauthoryear{Murtagh and
  Contreras}{2012}]{murtagh2012algorithms}
\begin{barticle}[author]
\bauthor{\bsnm{Murtagh},~\bfnm{Fionn}\binits{F.}} \AND
  \bauthor{\bsnm{Contreras},~\bfnm{Pedro}\binits{P.}}
(\byear{2012}).
\btitle{Algorithms for hierarchical clustering: an overview}.
\bjournal{Wiley Interdisciplinary Reviews: Data Mining and Knowledge Discovery}
\bvolume{2}
\bpages{86--97}.
\end{barticle}
\endbibitem

\bibitem[\protect\citeauthoryear{Ng et~al.}{2026}]{ng2026inference}
\begin{barticle}[author]
\bauthor{\bsnm{Ng},~\bfnm{Wai~Leong}\binits{W.~L.}},
  \bauthor{\bsnm{Tang},~\bfnm{Xinyi}\binits{X.}},
  \bauthor{\bsnm{Cheung},~\bfnm{Mun~Lau}\binits{M.~L.}},
  \bauthor{\bsnm{Gao},~\bfnm{Jiacheng}\binits{J.}},
  \bauthor{\bsnm{Yau},~\bfnm{Chun~Yip}\binits{C.~Y.}} \AND
  \bauthor{\bsnm{Dette},~\bfnm{Holger}\binits{H.}}
(\byear{2026}).
\btitle{Inference for Multiple Change-points in Piecewise Locally Stationary
  Time Series}.
\bjournal{arXiv preprint arXiv:2601.07400}.
\end{barticle}
\endbibitem

\bibitem[\protect\citeauthoryear{Ryan and Ulrich}{2022}]{quantmodpackage}
\begin{bmisc}[author]
\bauthor{\bsnm{Ryan},~\bfnm{Jeffrey~A.}\binits{J.~A.}} \AND
  \bauthor{\bsnm{Ulrich},~\bfnm{Joshua~M.}\binits{J.~M.}}
(\byear{2022}).
\btitle{{quantmod: Quantitative Financial Modelling Framework}}.
\bnote{R package version 0.4.20}.
\end{bmisc}
\endbibitem

\bibitem[\protect\citeauthoryear{Sun et~al.}{2024}]{sun2024multi}
\begin{barticle}[author]
\bauthor{\bsnm{Sun},~\bfnm{Yan}\binits{Y.}},
  \bauthor{\bsnm{Wan},~\bfnm{Chuang}\binits{C.}},
  \bauthor{\bsnm{Zhang},~\bfnm{Wenyang}\binits{W.}} \AND
  \bauthor{\bsnm{Zhong},~\bfnm{Wei}\binits{W.}}
(\byear{2024}).
\btitle{A Multi-Kink quantile regression model with common structure for panel
  data analysis}.
\bjournal{Journal of Econometrics}
\bvolume{239}
\bpages{105304}.
\end{barticle}
\endbibitem

\bibitem[\protect\citeauthoryear{Sun et~al.}{2025}]{sun2025homogeneity}
\begin{barticle}[author]
\bauthor{\bsnm{Sun},~\bfnm{Yan}\binits{Y.}},
  \bauthor{\bsnm{Tan},~\bfnm{Liming}\binits{L.}},
  \bauthor{\bsnm{Zhang},~\bfnm{Wenyang}\binits{W.}} \AND
  \bauthor{\bsnm{Zhu},~\bfnm{Zhenyu}\binits{Z.}}
(\byear{2025}).
\btitle{Homogeneity Pursuit in Clustered Data Analysis When Cluster Sizes Are
  Small}.
\bjournal{Journal of Business \& Economic Statistics}
\bpages{1--12}.
\end{barticle}
\endbibitem

\bibitem[\protect\citeauthoryear{Sundararajan and
  Pourahmadi}{2018}]{sundararajan2018nonparametric}
\begin{barticle}[author]
\bauthor{\bsnm{Sundararajan},~\bfnm{Raanju~R}\binits{R.~R.}} \AND
  \bauthor{\bsnm{Pourahmadi},~\bfnm{Mohsen}\binits{M.}}
(\byear{2018}).
\btitle{Nonparametric change point detection in multivariate piecewise
  stationary time series}.
\bjournal{Journal of Nonparametric Statistics}
\bvolume{30}
\bpages{926--956}.
\end{barticle}
\endbibitem

\bibitem[\protect\citeauthoryear{Tian and Feng}{2023}]{Tian02102023}
\begin{barticle}[author]
\bauthor{\bsnm{Tian},~\bfnm{Ye}\binits{Y.}} \AND
  \bauthor{\bsnm{Feng},~\bfnm{Yang}\binits{Y.}}
(\byear{2023}).
\btitle{Transfer Learning Under High-Dimensional Generalized Linear Models}.
\bjournal{Journal of the American Statistical Association}
\bvolume{118}
\bpages{2684--2697}.
\bdoi{10.1080/01621459.2022.2071278}
\end{barticle}
\endbibitem

\bibitem[\protect\citeauthoryear{Vogt and Linton}{2020}]{VOGT2020305}
\begin{barticle}[author]
\bauthor{\bsnm{Vogt},~\bfnm{Michael}\binits{M.}} \AND
  \bauthor{\bsnm{Linton},~\bfnm{Oliver}\binits{O.}}
(\byear{2020}).
\btitle{Multiscale clustering of nonparametric regression curves}.
\bjournal{Journal of Econometrics}
\bvolume{216}
\bpages{305--325}.
\end{barticle}
\endbibitem

\bibitem[\protect\citeauthoryear{Wand}{2021}]{kernsmoothpack}
\begin{bmisc}[author]
\bauthor{\bsnm{Wand},~\bfnm{Matt}\binits{M.}}
(\byear{2021}).
\btitle{{KernSmooth: Functions for Kernel Smoothing Supporting Wand \& Jones
  (1995)}}.
\bnote{R package version 2.23-20}.
\end{bmisc}
\endbibitem

\bibitem[\protect\citeauthoryear{Wang and Samworth}{2018}]{wangsamworth2018}
\begin{barticle}[author]
\bauthor{\bsnm{Wang},~\bfnm{Tengyao}\binits{T.}} \AND
  \bauthor{\bsnm{Samworth},~\bfnm{Richard~J.}\binits{R.~J.}}
(\byear{2018}).
\btitle{High dimensional change point estimation via sparse projection}.
\bjournal{Journal of the Royal Statistical Society: Series B (Statistical
  Methodology)}
\bvolume{80}
\bpages{57--83}.
\end{barticle}
\endbibitem

\bibitem[\protect\citeauthoryear{Wang and Samworth}{2020}]{inspectpackage}
\begin{bmisc}[author]
\bauthor{\bsnm{Wang},~\bfnm{Tengyao}\binits{T.}} \AND
  \bauthor{\bsnm{Samworth},~\bfnm{Richard}\binits{R.}}
(\byear{2020}).
\btitle{{InspectChangepoint: High-Dimensional Changepoint Estimation via Sparse
  Projection}}.
\bnote{R package version 1.1}.
\end{bmisc}
\endbibitem

\bibitem[\protect\citeauthoryear{Wang, Yu and Rinaldo}{2021}]{wang2021optimal}
\begin{barticle}[author]
\bauthor{\bsnm{Wang},~\bfnm{Daren}\binits{D.}},
  \bauthor{\bsnm{Yu},~\bfnm{Yi}\binits{Y.}} \AND
  \bauthor{\bsnm{Rinaldo},~\bfnm{Alessandro}\binits{A.}}
(\byear{2021}).
\btitle{Optimal change point detection and localization in sparse dynamic
  networks}.
\bjournal{The Annals of Statistics}
\bvolume{49}
\bpages{203--232}.
\end{barticle}
\endbibitem

\bibitem[\protect\citeauthoryear{Wang et~al.}{2022}]{2022wangvolgushev}
\begin{barticle}[author]
\bauthor{\bsnm{Wang},~\bfnm{Rui}\binits{R.}},
  \bauthor{\bsnm{Zhu},~\bfnm{Cheng}\binits{C.}},
  \bauthor{\bsnm{Volgushev},~\bfnm{Stanislav}\binits{S.}} \AND
  \bauthor{\bsnm{Shao},~\bfnm{Xiaofeng}\binits{X.}}
(\byear{2022}).
\btitle{Inference for change points in high-dimensional data via
  self-normalization}.
\bjournal{The Annals of Statistics}
\bvolume{50}
\bpages{781--806}.
\end{barticle}
\endbibitem

\bibitem[\protect\citeauthoryear{Wu and Zhou}{2018}]{wu2018gradient}
\begin{barticle}[author]
\bauthor{\bsnm{Wu},~\bfnm{Wei}\binits{W.}} \AND
  \bauthor{\bsnm{Zhou},~\bfnm{Zhou}\binits{Z.}}
(\byear{2018}).
\btitle{Gradient-based structural change detection for nonstationary time
  series M-estimation}.
\bjournal{The Annals of Statistics}
\bvolume{46}
\bpages{1197--1224}.
\end{barticle}
\endbibitem

\bibitem[\protect\citeauthoryear{Wu and Zhou}{2024}]{wu2019multiscale}
\begin{barticle}[author]
\bauthor{\bsnm{Wu},~\bfnm{Wei}\binits{W.}} \AND
  \bauthor{\bsnm{Zhou},~\bfnm{Zhou}\binits{Z.}}
(\byear{2024}).
\btitle{Multiscale jump testing and estimation under complex temporal
  dynamics}.
\bjournal{Bernoulli}
\bvolume{30}
\bpages{2372--2398}.
\end{barticle}
\endbibitem

\bibitem[\protect\citeauthoryear{Zhang, Wang and
  Shao}{2022}]{zhang2022adaptive}
\begin{barticle}[author]
\bauthor{\bsnm{Zhang},~\bfnm{Yifan}\binits{Y.}},
  \bauthor{\bsnm{Wang},~\bfnm{Rui}\binits{R.}} \AND
  \bauthor{\bsnm{Shao},~\bfnm{Xiaofeng}\binits{X.}}
(\byear{2022}).
\btitle{Adaptive inference for change points in high-dimensional data}.
\bjournal{Journal of the American Statistical Association}
\bvolume{117}
\bpages{1751--1762}.
\end{barticle}
\endbibitem

\bibitem[\protect\citeauthoryear{Zhou}{2013}]{zhou2013heteroscedasticity}
\begin{barticle}[author]
\bauthor{\bsnm{Zhou},~\bfnm{Zhou}\binits{Z.}}
(\byear{2013}).
\btitle{Heteroscedasticity and autocorrelation robust structural change
  detection}.
\bjournal{Journal of the American Statistical Association}
\bvolume{108}
\bpages{726--740}.
\end{barticle}
\endbibitem

\end{thebibliography}

\end{document}